\documentclass[%
 aip,
 amsmath,amssymb
 preprint,%
]{revtex4-1}

\usepackage{graphicx}
\usepackage{dcolumn}
\usepackage{bm}

\usepackage[electronic]{ifsym}
\usepackage[utf8]{inputenc}
\usepackage[T1]{fontenc}
\usepackage{times}
\usepackage{etoolbox}
\usepackage{subfigure}
\usepackage{multirow}

\usepackage{hyperref}
\hypersetup{
	colorlinks=true,
	linkcolor=blue,
	filecolor=blue,
	urlcolor=blue,
	citecolor=blue,
}
\usepackage{color}

  \usepackage{natbib}
\makeatletter
\def\@email#1#2{%
 \endgroup
 \patchcmd{\titleblock@produce}
  {\frontmatter@RRAPformat}
  {\frontmatter@RRAPformat{\produce@RRAP{*#1\href{mailto:#2}{#2}}}\frontmatter@RRAPformat}
  {}{}
}%
\makeatother
\begin{document}

\preprint{}

\title{Distinguishing thixotropy, anti-thixotropy, and viscoelasticity using hysteresis}%
\author{Yilin Wang}
\affiliation{ 
	Department of Mechanical Science and Engineering, University of Illinois Urbana-Champaign, Urbana, IL 61801, USA}
\affiliation{ 
	Beckman Institute for Advanced Science and Technology, University of Illinois Urbana-Champaign, Urbana, IL 61801, USA}
\affiliation{ 
	Joint Center for Energy Storage Research, Argonne National Laboratory, Lemont, IL 60439, USA
}
\author{Randy H. Ewoldt}
  \altaffiliation[Author to whom correspondence should be addressed;]{ electronic email: ewoldt@illinois.edu}
\affiliation{ 
	Department of Mechanical Science and Engineering, University of Illinois Urbana-Champaign, Urbana, IL 61801, USA}
\affiliation{ 
	Beckman Institute for Advanced Science and Technology, University of Illinois Urbana-Champaign, Urbana, IL 61801, USA}
\affiliation{ 
	Joint Center for Energy Storage Research, Argonne National Laboratory, Lemont, IL 60439, USA
}

\date{\today}

\begin{abstract}
Thixotropy, anti-thixotropy, and viscoelasticity are three types of time-dependent dynamics that involve fundamentally different underlying physical processes. Yet distinguishing them can be very challenging, which hinders the understanding of structure-property relations. Here we show that hysteresis is a promising technique to contrast the three dynamics by exploring signatures of the most basic thixotropic, anti-thixotropic, and nonlinear viscoelastic models. From these signatures, using shear-rate controlled ramps that begin and end at high shear rates, we identify two distinguishing features in hysteresis loops. The first is the direction of the hysteresis loops: clockwise for thixotropy, but counterclockwise for viscoelasticity and anti-thixotropy. A second feature is achieved at high ramping rates where all responses lose hysteresis: the viscoelastic response shows a stress plateau at low shear rates (lack of stress relaxation), whereas the thixotropic and anti-thixotropic responses are purely viscous with minimal shear thinning or thickening. The features are observed independent of the model details. We establish further evidence for these signatures by experimentally measuring the hysteresis of thixotropic Laponite suspensions, anti-thixotropic carbon black suspensions, and viscoelastic poly (ethylene oxide) solutions. The protocols explored here can be used to distinguish thixotropy, anti-thixotropy, and viscoelasticity, which helps reveal the underlying microstructural physics of complex fluids.

\end{abstract}

\maketitle


\section{\label{sec:introduction}Introduction}
Many complex fluids exhibit flow-sensitive and time-dependent rheological properties, such as thixotropy, anti-thixotropy, and viscoelasticity. For example, most attractive colloids can form a space-filling network in suspension at volume fractions higher than the percolation threshold \cite{DaSilvaLeiteCoelho2017, Richards2017}, thus demonstrating viscoelastic properties \cite{Trappe2000, Trappe2001}. The network can break down in time under applied shear stress that is large enough, resulting in thixotropy \cite{Usersbook}. Under certain conditions, the structural rearrangement \cite{Hipp2019, Hipp2021}, such as interpenetration and densification \cite{Ovarlez2013}, leads to anti-thixotropy \cite{Wang2022, Narayanan2017}. Thixotropy, anti-thixotropy, and viscoelasticity can result simultaneously in the same system. Thus, distinguishing the three dynamics is important to understand the underlying microstructural physics and to design complex fluids with specifically desired properties for applications \cite{Nelson2017, Ewoldt2022}, such as 3D printing \cite{Ewoldt2022, Wang2008, Corker2019, Smay2002}, semi-solid flow batteries \cite{Youssry2013, Duduta2011, Fan2014,Wei2015}, and coating \cite{Sen2021}.  

Thixotropic fluids are common both in industry and in daily life, including the slurry in flow batteries \cite{Yearsley2012, Youssry2013}, human blood \cite{Horner2019}, mud/kaolinite clay suspensions \cite{Ran2023}, fumed silica suspensions \cite{ Dullaert2005c}, carbon black suspensions \cite{Dullaert2005a, Dullaert2005b, Dullaert2006, Wang2022}, crude oil \cite{Chang1999}, and bentonite-water mixture \cite{Toorman1997}. The word thixotropy has been used broadly in literature to refer to a time-dependent, isothermal, and reversible change in a material property, such as the elastic modulus, yield stress, or shear viscosity, wherein the flow resistance decreases under shearing (also called rejuvenation), and increases again when the flow strength is decreased (also called aging) \cite{Wang2022, Sen2021, Ovarlez2013, Bauer1967}. For example, IUPAC defines thixotropy as "the continuous decrease of viscosity with time when flow is applied to a sample that has been previously at rest, and the subsequent recovery of viscosity when flow is discontinued \cite{IUPAC1997, Mewis2009}." We personally find it unnecessarily limiting to restrict the term thixotropy to apply only to observed viscosity, since in a real system, other properties, such as moduli, will also change in time due to the same underlying structural evolution. As such, here we use thixotropy and anti-thixotropy in the broader sense to refer to time-dependent property change during shear.

\begin{figure}[hbt!]
	\centering
	\includegraphics[width=0.65\textwidth]{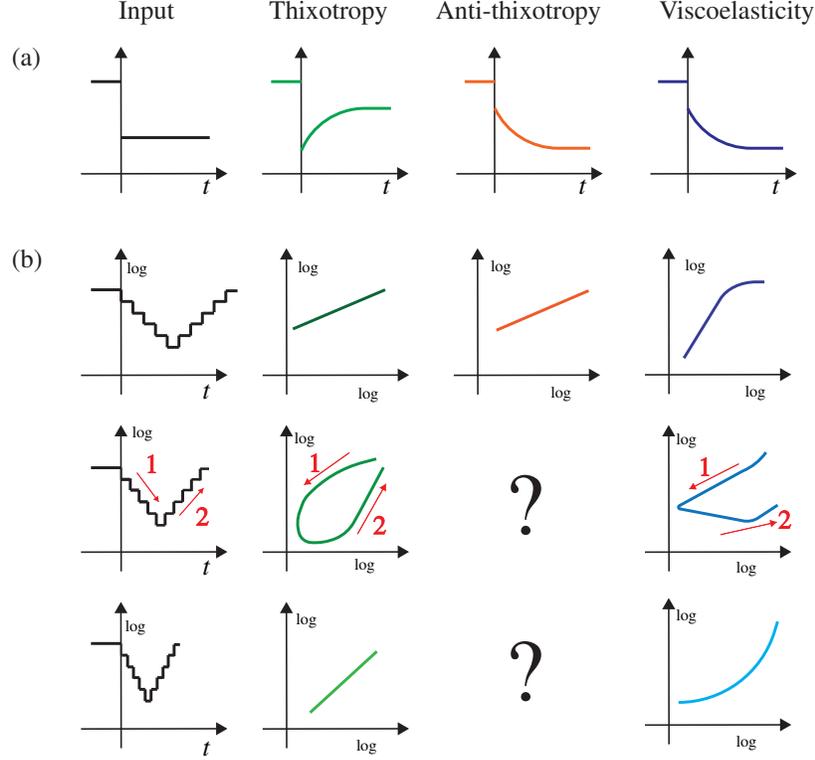}
	\hspace{0.01\linewidth}
	\caption{Simple step in shear rate cannot differentiate between anti-thixotropy and viscoelasticity, though it is commonly used to differentiate between thixotropy and viscoelasticity, as shown in (a) the schematics of transient shear stress evolution with time in response to a step-down in shear rate flow, the anti-thixotropy and viscoelasticity show the same stress response trend. (b) Hysteresis loop in response to ramping-down and up in shear rate tests can potentially help distinguish thixotropy, anti-thixotropy, and viscoelasticity, which we explore. Some features of hysteresis have been reported: the area of thixotropic hysteresis shows a ramping time dependence, first increasing then decreasing with the duration \cite{Divoux2013}; the viscoelastic hysteresis shows shear rate dependence and self-intersection can appear \cite{Sharma2022}; for anti-thixotropic materials, the features of the loop are unknown.}
	\label{Fig.schematics}
\end{figure}

A common protocol to demonstrate thixotropy is the step-down shear rate test, as shown in Fig.~\ref{Fig.schematics}\textcolor{blue}{(a)}. When the applied shear rate is decreased from $\dot \gamma_i$ to $\dot \gamma_f$, the shear stress for a thixotropic fluid shows a sudden decrease, followed by a gradual increase until steady state is reached, because of the increase of flow resistance under lower applied shear rate. Recently, it was shown that from this stress response, a thixotropic spectra can be defined to quantify and compare thixotropy, that is, a superposition of essential stress modes distributed over thixotropic timescales \cite{Sen2022}. This acknowledges a range of underlying timescales and enables the quantification of average and dispersity of timescales for both build-up and break-down processes. 

A special form of thixotropy, anti-thixotropy, acts in the opposite direction to thixotropy. Anti-thixotropy is also called rheopexy, or negative thixotropy, and its definition can be extended from thixotropy \cite{Larson2015}, that is, an isothermal and reversible change in material properties, so that the flow resistance increases under shearing, and decreases when the applied shear rate is decreased. As shown in Fig.~\ref{Fig.schematics}\textcolor{blue}{(a)}, for an anti-thixotropic fluid, the stress decreases under a step-down in shear rate, because of a decrease in flow resistance under lower shear rate (this is not associated with viscoelastic stress relaxation, and it can appear even in purely viscous anti-thixotropic fluids). Some studies use the term "shear-induced overaging", where a small applied deformation or shear rate increases the rate of aging of viscoelastic properties compared to the viscoelastic aging at zero-stress conditions \cite{Sudreau2022, Agarwal2020, Viasnoff2002}. This is similar to anti-thixotropy, but here we use anti-thixotropy to refer to material evolution under flow, e.g., imposed shear rate, that is fundamentally different than what would occur at zero stress conditions. The transient stress response of anti-thixotropy has been observed in different types of suspensions such as carbon black \cite{Wang2022}, coal-water \cite{Keller1991}, and nuclear waste simulant slurries \cite{Chang1996}. It can happen with certain attractive particles where shear promotes temporary aggregation rather than breaking the structure \cite{Chang1996}, e.g. due to collision of particles. It can also arise as a result of shear-induced densification \cite{Vermant2005, Hoekstra2003, Varadan2001, Wang2022}.

Viscoelasticity is associated with time-dependence due to elastic stress storage and subsequent relaxation, as directly revealed by the relaxation modulus material function, $G(t)$, decreasing with time. As shown in Fig.~\ref{Fig.schematics}\textcolor{blue}{(a)}, stress in a viscoelastic material shows a continuous decrease in the transient shear stress response in step-down in shear rate tests. A step-down (or step-up) in shear rate test is a common experiment to distinguish thixotropy from viscoelastic stress relaxation \cite{Mewis2009}, and other protocols proposed to discriminate between thixotropy and viscoelasticity include applying step-strain or step-stress after the cessation of preshear \cite{Agarwal2021}. However, all the protocols mentioned above fail to distinguish anti-thixotropy from thixotropy and viscoelasticity. On the other hand, viscoelastic materials showing nonlinear effects such as shear-thinning can show features that resemble that of thixotropy. For example, the stress undershoot (or overshoot in case of step-up in shear rate and startup flow) for nonlinear viscoelasticity \cite{Saengow2019} shows a stress increase with time, which is indistinguishable from stress recovery due to thixotropic structural change.  

To differentiate thixotropy, anti-thixotropy, and viscoelasticity, Wang and Ewoldt \cite{Wang2022} used orthogonal superposition (OSP) \cite{Vermant1998}, where a thixotropic system shows an increase in orthogonal moduli in step down shear rate flow, an anti-thixotropic system shows a decrease, whereas a viscoelastic system shows a constant orthogonal moduli (linear viscoelasticity) or an increase (nonlinear viscoelasticity showing shear thinning behavior). Therefore, together with step down tests, OSP can help differentiate the three dynamics. However, the evolution of orthogonal moduli has only been studied for a few constitutive models and material systems \cite{Kim2013, Zhang2021}. For more general thixo-elasto-visco-plastic (TEVP) materials and models \cite{Varchanis2019, Ewoldt2017}, the trend of change in orthogonal moduli is not clear yet. Another caveat of using OSP is the complicated applied flow field, which can significantly affect the properties of fluids \cite{Lin2016}, so that the measured orthogonal moduli may not represent the dynamics of the system without the superposed oscillation. All these factors make distinguishing thixotropy, anti-thixotropy, and viscoelasticity very challenging.

We propose to use hysteresis to distinguish the three dynamics. Using hysteresis to study thixotropy was first introduced by Green and Weltmann \cite{Green1943, Green1944}. Although it is less preferred than the step in shear rate test because in hysteresis, the effect of shear rate and time cannot be separated, it is still a commonly used way to demonstrate thixotropy. Hysteresis loops have been observed on a number of thixotropic systems \cite{Green1944, Mewis2009, Divoux2013, Toorman1997}, and the degree of thixotropy is related to the area of a  hysteresis loop: the larger the loop, the more thixotropic the response is \cite{Green1943}. The area of hysteresis loops has been proved to be ramping time dependent. As shown in Fig.~\ref{Fig.schematics}\textcolor{blue}{(b)}, when the ramping time is large compared to the thixotropic timescale, the downward and upward ramping curves overlap and there is no hysteresis loop. The area of a hysteresis loop is maximized when the ramping time is comparable to the dominant (average) thixotropic time of the system, and the hysteresis gradually decreases to zero as the ramping time decreases. The hysteresis for anti-thixotropy is less studied. It has been observed on a couple of anti-thixotropic systems, such as carbon black suspensions \cite{Negi2009, Wang2022} and aqueous xanthan gum solutions \cite{Ngouamba2021}. Viscoelastic hysteresis was first studied by Marsh and Bird \cite{Bird1968, Marsh1968}, where several types of viscoelastic hysteresis loops were obtained for the Bird-Carreau model and four kinds of nonlinear viscoelastic polymer solutions. Hysteresis for different linear and nonlinear viscoelastic models has been studied recently. For example, hysteresis loops for single mode Maxwell model \cite{Rubio2008} were calculated under different ramping times; hysteresis for the Giesekus model was first studied by Wang and Ewoldt \cite{Wang2021}, and the shapes of hysteresis loops for the Giesekus model were then calculated by Sharma and coworkers \cite{Sharma2022}. For both linear and nonlinear models, the area of hysteresis loops shows ramping time dependence and significant loops are observable at intermediate times, as shown in Fig.~\ref{Fig.schematics}\textcolor{blue}{(b)}. However, a clear comparison of fingerprints of thixotropy, anti-thixotropy, and viscoelasticity in hysteresis has not been done. Therefore, in this study, we numerically and experimentally study the fingerprints of the three dynamics in hysteresis, and identify different fingerprints, such as the direction of the loop, the shape of the loops, and the self-intersection, and we use those distinguishing features to differentiate between thixotropy, anti-thixotropy, and viscoelasticity.

In this paper, we numerically calculate the hysteresis loops of the most fundamental thixotropic kinetic model \cite{Larson2015}, an anti-thixotropic kinetic model (newly constructed), and the viscoelastic Giesekus model \cite{DPL1987} under different ramping times and applied shear rate ranges. We also experimentally identify and study three materials, a thixotropic Laponite suspension, an anti-thixotropic carbon black suspension, and a nonlinear viscoelastic PEO solution, and we measure the hysteresis loops of those three materials. The experimental results observed are consistent with the model predictions. There are different ways to generate hysteresis. In this paper, a rate-controlled, discrete down and up ramping is used, which is explained in Sec. \ref{sec:hysteresis}. To compare the fingerprints of hysteresis loops, the most fundamental thixotropic and anti-thixotropic models are used here. The constitutive models and governing equations are introduced in Sec. \ref{sec:models}, and the hysteresis loops predicted by the models are shown in Sec. \ref{sec:results_model}. The experimentally measured hysteresis loops are in Sec. \ref{sec:results_exp}.

\section{\label{sec:materialandmethod}Materials and methods}

\subsection{\label{sec:hysteresis}Hysteresis protocol}

To generate hysteresis, the shear rate is ramped down from a maximum value, $\dot \gamma_{\rm max}$, to a minimum, $\dot \gamma_{\rm min}$, through $n$ logarithmically spaced steps per decade with $\delta t$ duration per step, then ramped up following the same steps. Here, $n$ is the number of steps per decade of shear rate, and $\delta t$ is the duration per step. Therefore, the ramping time per decade is $T=n \delta t$. In this study, before hysteresis ramping, preshear at the the maximum shear rate is applied to erase all mechanical history \cite{Jamali2022}. Starting a hysteresis loop at $\dot \gamma_{\rm max}$ ensures that the system is starting from equilibrium so that the self-intersection in hysteresis loops can be minimized.

\begin{figure}[h!]
	\centering
	\subfigure[]{
		\label{Fig.Protocol_thixo_T1_n2_rate_time}
		\includegraphics[width=0.3\textwidth]{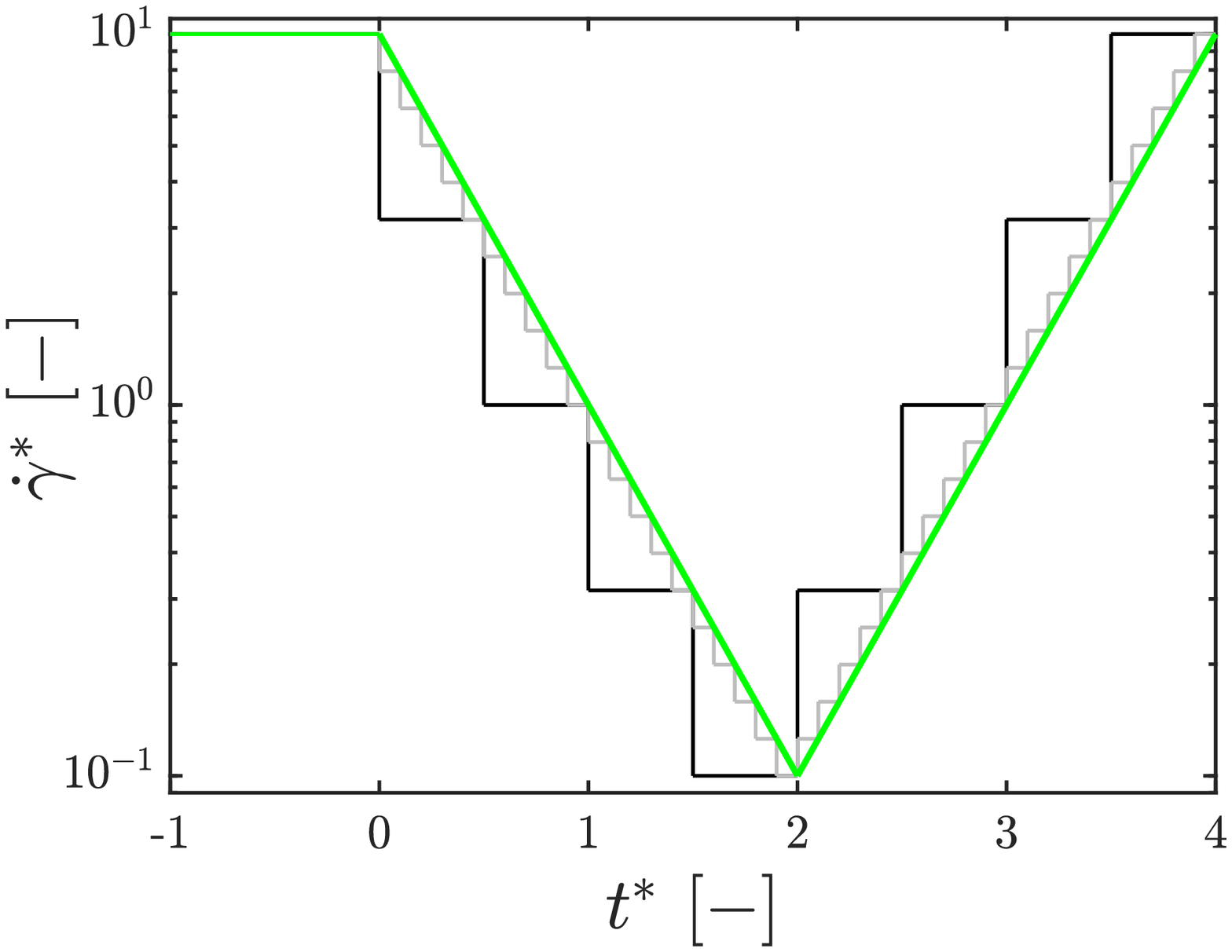}}
	\subfigure[]{
		\label{Fig.Protocol_thixo_T1_n2_stress_time}
		\includegraphics[width=0.3\textwidth]{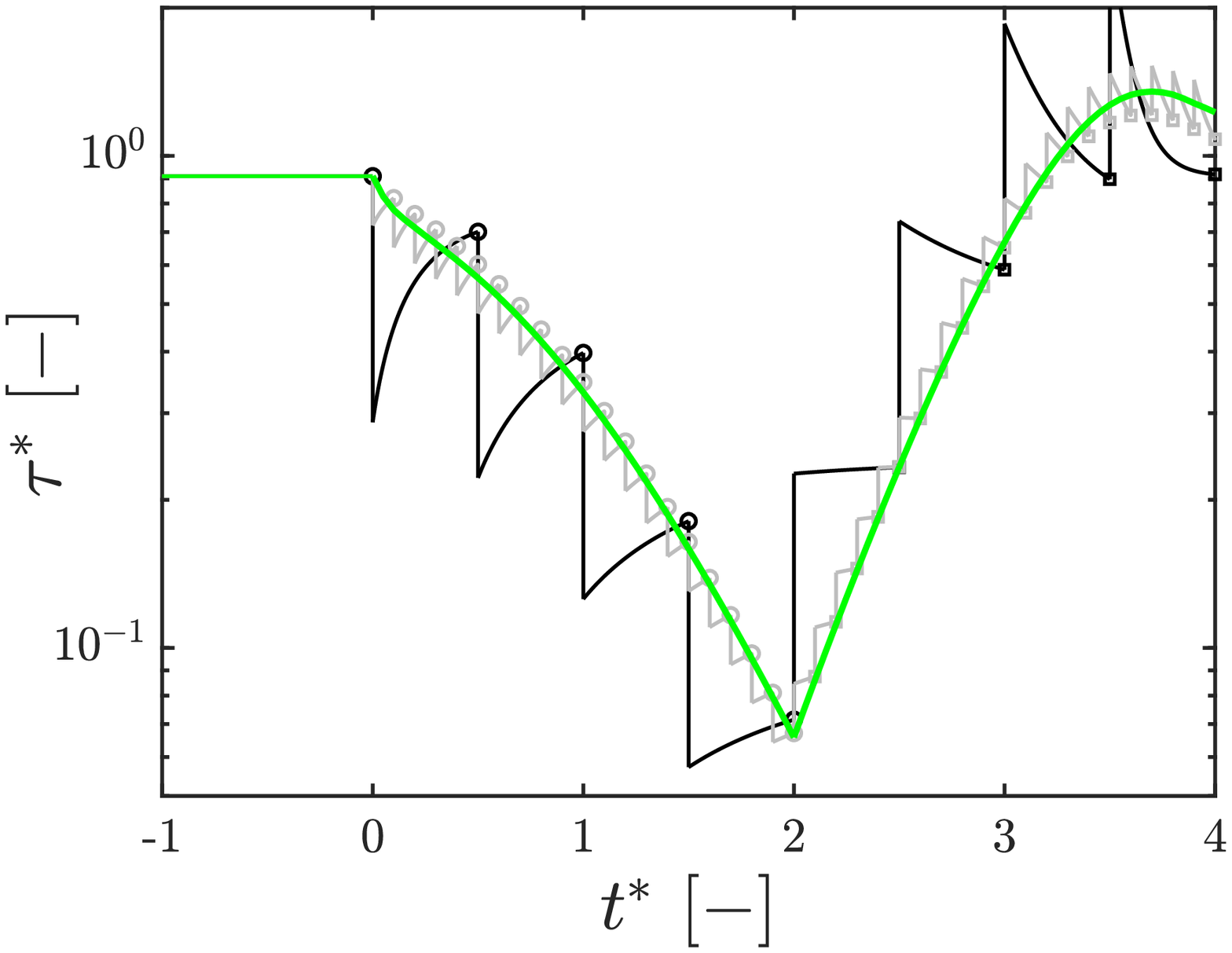}}
	\hspace{0.01\linewidth}
		\subfigure[]{
		\label{Fig.Protocol_thixo_T1_n2_stress_rate}
		\includegraphics[width=0.3\textwidth]{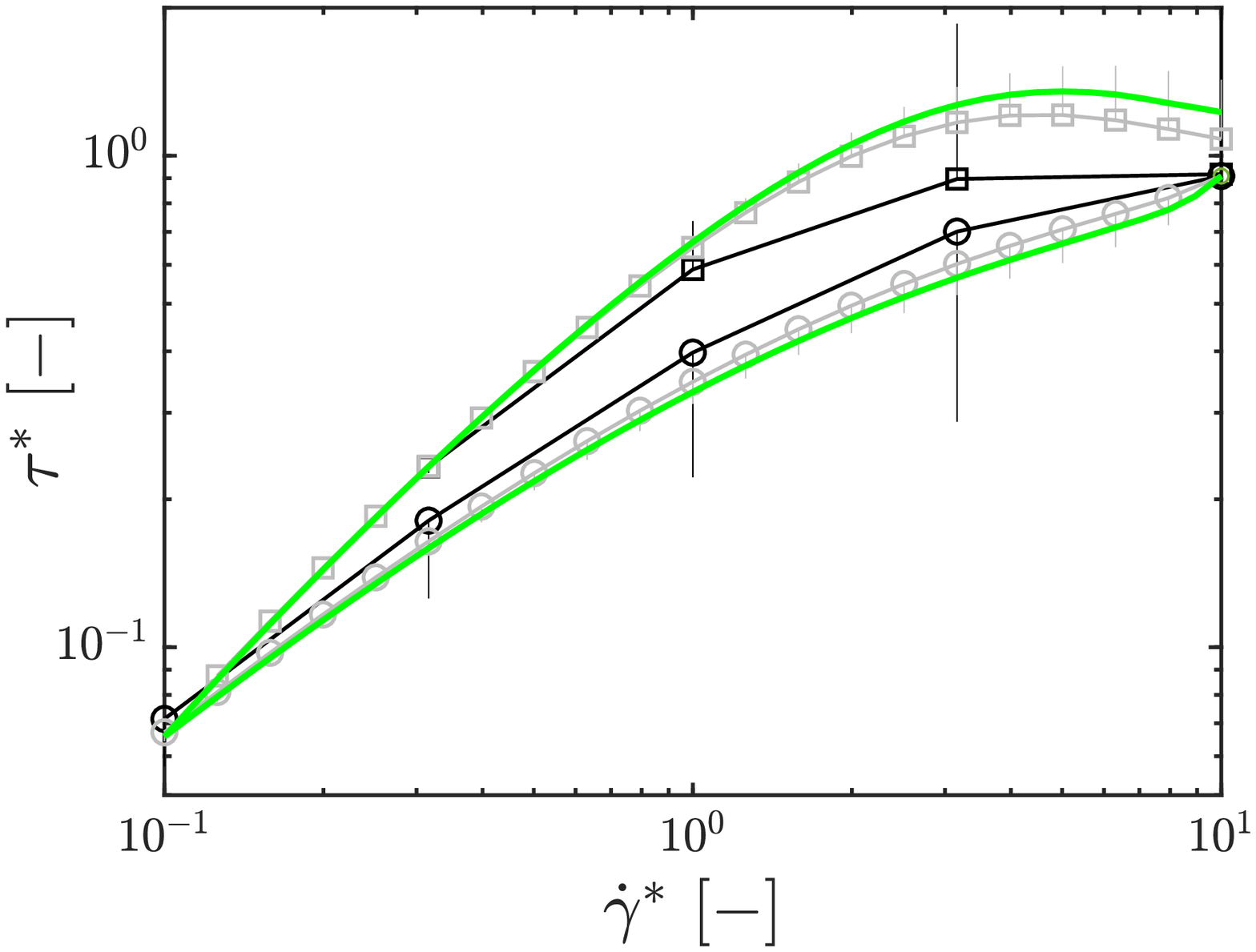}}
	\hspace{0.01\linewidth}
	\caption{Discrete versus continuous ramp protocol, and consequences for a simple dimensionless thixotropic structure-parameter model (Eq.~\ref{Eq.thixo_stress_dimensionless} and \ref{Eq.thixo_structure_dimensionless}) introduced in Sec.~\ref{sec:thixo_model}. (a) Input shear scheduling protocol for $n=2$, $\delta t^* = 0.5$ (black), $n=10$, $\delta t^* = 0.1$ (grey), and continuous shear input (green), the total time per decade is $T^* =n \delta t^* =1$; before time zero, the shear rate is kept at the maximum value, which is 10 in this case. (b) The resulting stress response schematics for thixotropy: at each shear rate, the stress increases with time and the stress at the end of each step is chosen to be the stress at the shear rate. (c) Processing the data into hysteresis loop, the open circle shows the stress at the end of each step in the downward ramping and the open square shows those in the upward ramping, the vertical lines extending below and above the data show the transient stress during the ramping.}
	\label{Fig.protocol}
\end{figure}

Both discrete and continuous ramp protocols can be used to generate hysteresis. Discrete ramping is easier to implement but exhibits stronger protocol dependence, that is, the hysteresis loop depends on the number of points per decade, $n$, even if the ramping time per decade of shear rate, $T$, is kept the same. In this study, a discrete ramp protocol with a large enough number of points is used to approach continuous ramping. As shown in Fig.~\ref{Fig.protocol}\textcolor{blue}{(a)}, the black line represents the protocol for $n=2$, $\delta t^* = 0.5$, and the grey line is for $n=10$, $\delta t^* = 0.1$. We use $^*$ to represent dimensionless variables, and $\delta t^*$ is the dimensionless duration defined as $\delta t^* = \frac{\delta t}{t_{\rm char}}$, where $t_{\rm char}$ is the characterized timescale of specific model. Therefore, the ramping time per decade of shear rate, $T^*= n \delta t^* = 1$, for each case. The green line represents the continuous ramping with the ramping time per decade of shear rate $T^*=1$. Detailed nondimensionalization and the characterized shear rate of each model will be introduced in Sec.~\ref{sec:models}. For models, we use a dimensionless shear rate range from $\dot \gamma_{\rm max}^* = 10$ to $\dot \gamma_{\rm min}^* = 0.1$. and the dimensionless shear rate is defined as $\dot \gamma^* = \frac{\dot \gamma}{\dot \gamma_{\rm char}}$, where $\dot \gamma_{\rm char}$ is the characteristic shear rate of the model. For experiments, different ranges of shear rate are used for different materials tested, shown in Sec.~\ref{sec:results_exp}. The consequences of the discrete and continuous ramping protocols are shwon in Fig.~\ref{Fig.Protocol_thixo_T1_n2_stress_time}-\ref{Fig.Protocol_thixo_T1_n2_stress_rate} using a simple dimensionless thixotropic structure-parameter model (details in Sec.~\ref{sec:thixo_model} Eqs.~\ref{Eq.thixo_stress_dimensionless} and \ref{Eq.thixo_structure_dimensionless}).

We calculate the evolution of dimensionless shear stress, $\tau^*$, with dimensionless time, $t^*$, during the ramping in Fig.~\ref{Fig.Protocol_thixo_T1_n2_stress_time}. Qualitatively, during the downward ramping, at each shear rate, the shear stress shows a sudden decrease followed by a gradual increase with time, because of the thixotropic structure build-up with the decrease in shear rate. The opposite trend for shear stress can be observed in the upward ramping. The stress at the end of duration for each shear rate is recorded and plotted as a function of applied shear rate, generating a hysteresis loop, as shown in Fig.~\ref{Fig.Protocol_thixo_T1_n2_stress_rate}. For discrete ramping, as shown in black and grey, the range of transient stress at each shear rate is plotted as vertical lines. It can be seen from the plot that the black loop, where the number per decade $n=2$, has a larger transient and deviates more from the continuous ramping plotted in green. For the same ramping time, $T^*=1$, we calculate the deviation of hysteresis loops to the continuous ramping and plot it as a function of number per decade, $n$, as shown in the Supporting Information, Fig.~\textcolor{blue}{S4(b)}. The larger the $n$, the smaller the deviation; when $n$ is larger than 10, the average deviation is lower than 10$\%$ for all ramping times studied here. This also applies to other ramping times, as shown in Fig.~\textcolor{blue}{S2-S4}. Therefore, $n=10$ is used in this study for both numerical model calculations and experiments to generate hysteresis.

\subsection{\label{sec:method}Numerical and Experimental Details}
The stress evolution during ramping was solved analytically for the thixotropic and anti-thixotropic models to generate hysteresis, while for the nonlinear viscoelastic Giesekus model, the stress was solved with numerical integration using the ode45 function by MATLAB \cite{matlab}. Homogeneous simple shear flow is assumed for all models and fluid inertia is neglected for this transient flow. The governing equations for different models will be discussed in Sec.~\ref{sec:models}. 

Rheological measurements were performed at 25$\rm ^oC$ on a rate-controlled, separate motor-transducer rheometer (ARES-G2 from TA instruments) with a cone-and-plate geometry with a diameter of 25~mm and a cone angle of 0.1~rad.

\subsection{\label{sec:materials}Materials}

Three materials are used in this study: Laponite aqueous suspension,  carbon black (CB) suspension in heavy mineral oil, and aqueous solution of poly (ethylene oxide) (PEO). 

Laponite forms a soft gel when dispersed in aqueous solution. The 3~wt$\%$ Laponite suspensions used in this study were prepared by dissolving the required amounts of Laponite power (Conservation Support Systems, USA) in distilled water. The sample was sealed and left to stir using a magnetic stirrer for three days to achieve a homogeneous and stable suspension. 

The CB used in this study is Vulcan XC-72 (Cabot Corporation), which has a density of $1,800$~kg/m$^3$. The CB particles were dispersed in a heavy mineral oil (Fisher Scientific), which has a density of $1,200$~kg/m$^3$ and a Newtonian viscosity of $0.14$~Pa$\cdot$s at 25$\rm ^oC$. The 8~wt$\%$ suspension was prepared using a rotor-stator homogenizer (IKA) by shearing at 4000~rpm for more than 10~min to make a well dispersed and stable CB suspension in mineral oil. 

PEO of nominal molecular weight of $8 \cdot 10^6$~g/mol was purchased from Sigma-Aldrich. The 1~wt$\%$ aqueous solution of PEO was prepared by dissolving the appropriate amount of PEO in co-solvent of 95$\%$ distilled water and 5$\%$ ethanol at room temperature. Sufficient time (24~hr) of continuous magnetic stirring was allowed to achieve complete homogenization. 

\section{\label{sec:models}Models and Governing Equations}

\begin{figure}[h!]
	\centering
		\includegraphics[width=0.95\textwidth]{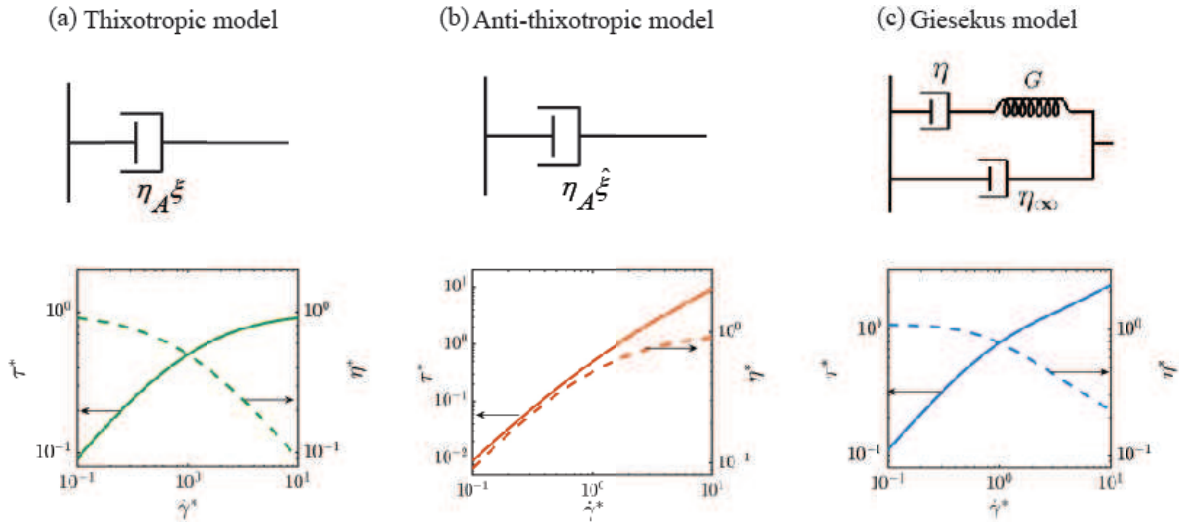}
	\hspace{0.01\linewidth}
	\caption{Mechanical analogy and the steady state stress and viscosity for (a) the thixotropic kinetic model (shown in green), (b) the anti-thixotropic kinetic model (orange), and (c) the Giesekus model (blue). The solid lines show the steady state shear stress, and the dashed lines show the shear viscosity. }
    \label{Fig.models}
\end{figure}

To calculate hysteresis loops for thixotropy, anti-thixotropy, and viscoelasticity, three fundamental models are used here. The mechanical analogy schematics and the steady state flow curves of the three models are shown in Fig.~\ref{Fig.models}. The details of each model will be introduced in this section separately, describing their governing differential equations in dimensional and dimensionless forms and showing the stress response in step-down shear rate tests.

\subsection{\label{sec:thixo_model}The simplest thixotropic kinetic model}

Thixotropic kinetic models usually couple a constitutive equation with a scalar variable which tracks the degree of "structure" of fluids. The structure parameter, $\xi$, can be interpreted as a conceptual measure of the degree of aggregation, bonding, or jamming. (The variable $\lambda$ is commonly used for this parameter in literature, however we reserve $\lambda$ for viscoelastic relaxation time.) The basic form of thixotropic kinetic equations describing the change of thixotropic structure was initially proposed by Goodeve and Whitfield \cite{Goodeve1938}, which is also called Moore's kinetic structure equation \cite{Moore1959} as
\begin{equation}
    \frac{d\xi}{d t} = k_A (1-\xi)-k_D \dot \gamma \xi,
\label{Eq.thixo_kinetics}
\end{equation}
where $\xi$ is dimensionless and ranges from zero to one. In Eq.~(\ref{Eq.thixo_kinetics}), the term $k_A (1-\xi)$ accounts for the structural formation, e.g. assisted by thermal fluctuations, and $k_D \dot \gamma \xi$ is the structural breakage due to shear, which is proportional to the shear rate, $\dot \gamma$. The two kinetic coefficients, $k_A$ and $k_D$, are constants. Many thixotropic models use more complex kinetic equations, adding another term, using multiple structure parameters \cite{Wei2016, Wei2018, Wei2019}, or including a power function exponent on the existing terms. To formulate the "simplest thixotropic model" \cite{Blackwell2014}, here we examine Eq.~(\ref{Eq.thixo_kinetics}) as the simplest first-order kinetic equation. This model is well known, and in the rest of this section, we take time with the details of this model to be clear about the dimensionless quantities, thixotropic build-up and break-down timescales, and the general response to step change in shear rate with arbitrary initial conditions. 

At steady state in Eq.~(\ref{Eq.thixo_kinetics}), the value of the structure parameter, $\xi$, depends on the shear rate, as 
\begin{equation}
    \xi_{\rm ss} = \frac{k_A}{k_A + k_D \dot \gamma} = \frac{1}{1+\frac{k_D}{k_A} \dot \gamma}. 
    \label{Eq.thixo_kinetics_dimensional_ss}
\end{equation}
This suggests a characteristic shear rate $\dot \gamma_{\rm char} = \frac{k_A}{k_D}$, which gives a dimensionless shear rate $\dot \gamma^*$:
\begin{equation}
    \dot \gamma^* = \frac{\dot \gamma}{\dot \gamma_{\rm char}} = \frac{k_D}{k_A} \dot \gamma.
\end{equation}
As (dimensionless) shear rate goes to zero, the structure parameter goes to one and the fluid is "fully structured"; as (dimensionless) shear rate goes to infinity, the structure parameter goes to zero and the fluid is "fully unstructured/broken down". 

The simplest constitutive model we use in this study is formed by coupling the structure parameter with a generalized Newtonian fluid, as shown in Fig.~\ref{Fig.models}. This model has only one viscous component, and we assume the viscosity, $\eta$, depends linearly on the thixotropic structure parameter, $\eta = \eta_A \xi$. Therefore, the stress, $\tau$, is, 
\begin{equation}
    \tau = \eta(\xi) \dot \gamma = \eta_A \xi \dot\gamma. 
\end{equation}
With the recovery of $\xi$ at low shear rates, $\tau$ also shows a recovery with time. One can form more complicated thixotropic models by adopting more complex stress-structure coupling equations, e.g. forming a thixo-elasto-visco-plastic (TEVP) model \cite{Varchanis2019, Ewoldt2017}. Here, we do not consider elasticity and plasticity for simplicity. A characteristic stress can be formed by $\tau_{\rm char} = \eta_A \dot \gamma_{\rm char} = \eta_A \frac{k_D}{k_A}$ and therefore the stress is nondimensionalized as 
\begin{equation}
    \tau^* = \frac{\tau}{\tau_{\rm char}} = \frac{k_D}{\eta_A k_A} \tau. 
\end{equation}

The structure-based model contains two timescales: the destruction timescale $\left( k_D \dot \gamma \right)^{-1}$ and the aggregation timescale $k_A^{-1}$. The effective thixotropic evolution timescale depends on both, as the change of structure is the result of competition between shear destruction and aggregation. In general, thixotropic evolution has two timescales to consider \cite{Sen2022, Sen2021}, the thixotropic build-up timescale, $\lambda_{\rm Thixo,+}$, and the break-down timescale,  $\lambda_{\rm Thixo,-}$. This is unlike linear viscoelasticity, where one timescale, $\lambda_{\rm VE}$, governs both elastic stress growth (e.g. $\eta^+$ in start-up flow) and stress relaxation (e.g. $G(t)$ in step strain). The build-up and break-down timescales are different in general because of shear history dependence, that is, depending on whether the stress is recovering from a higher initial shear rate, or decaying from a lower one, the recovery and decay timescales may be different even for the same final shear rate. Therefore, it is important to define two dimensionless timescales \cite{Sen2021}, 
\begin{equation}
    \mathcal{T}_+ = \frac{\rm observation~(experiment)~time}{\rm thixotropic~buildup~time} = \frac{t_{\rm exp}}{\lambda_{\rm Thixo,+}}
\end{equation}
and 
\begin{equation}
    \mathcal{T}_- = \frac{\rm observation~(experiment)~time}{\rm thixotropic~breakdown~time} = \frac{t_{\rm exp}}{\lambda_{\rm Thixo,-}}.
\end{equation}
During recovery tests (shear rate ramping down, aging), $\mathcal{T}_+$ quantifies the degree of structure build up. When $\mathcal{T}_+ \ll 1$, the structure does not have enough time to recover and remains at its initial state; when $\mathcal{T}_+ \sim 1$, the experiment time is comparable to the thixotropic recovery time and the structure recovers significantly; as $\mathcal{T}_+ \gg 1$, structure approaches its steady state. Similarly, for decay tests (shear rate ramping up, rejuvenation), $\mathcal{T}_-$ quantifies the degree of structure break down. One could also define thixotropic Deborah numbers \cite{Ewoldt2017} as $\rm De_{thixo,+} = 1/\mathcal{T}_+ = \frac{\lambda_{\rm Thixo,+}}{t_{\rm exp}}$ and $\rm De_{thixo,-} = 1/\mathcal{T}_-$, or interpret as a type of mutation number \cite{Jamali2022}, also called thixoviscous number in previous literature \cite{Ewoldt2017}. We find using $\mathcal{T}_+$ and $\mathcal{T}_-$ most useful and interpretable in this study as we vary the experiment ramping time for a model with its specified thixotropic build-up and break-down timescales. 

We can analytically derive the build-up and break-down timescales for the thixotropic kinetic equation used in this study. Subjecting Eq.~(\ref{Eq.thixo_kinetics}) to a step change in shear rate with an arbitrary initial condition, it is found that the structure parameter evolves exponentially with a characteristic timescale dependent on the new shear rate, and the build-up and break-down timescales are the same,
\begin{equation}
    \lambda_{\rm Thixo,+} = \lambda_{\rm Thixo,-} = \frac{1}{k_A + k_D \dot\gamma}.
\end{equation}

In the hysteresis protocol, the experiment timescales for ramping down and ramping up are also the same, which is the ramping time per decade of shear rate, $T$. In this case, $\mathcal{T}_+ = \mathcal{T}_- = \frac{T}{ \left( k_A + k_D \dot\gamma \right)^{-1}}$. This dimensionless time still depends on the applied shear rate, which is not a constant in hysteresis. For simplicity, we define a nominal dimensionless timescale, which is evaluated at zero shear rate, that is,
\begin{equation}
    T^* = \mathcal{T}_+(\dot \gamma=0) =  \mathcal{T}_-(\dot \gamma=0) = \frac{T}{\lambda_{\rm Thixo,\dot \gamma=0}} = \frac{T}{k_A^{-1}}  = k_A T.
    \label{Eq.thixo_define_timescale}
\end{equation}
Dimensionless time is defined in the same way: $t^* = k_A t$. It should be noted that $T^*$ defined in this way is an underestimate of $\mathcal{T}_+$ and $\mathcal{T}_-$, since the true thixotropic timescale is always shorter than $ k_A^{-1}$ when the shear rate is finite. The ratio between them is:
\begin{equation}
    \frac{T^*}{\mathcal{T}_+} = \frac{k_A}{k_A + k_D \dot\gamma} = \frac{1}{1+ \dot \gamma^*}.
    \label{Eq.thixo_timescale_definition}
\end{equation}
For the dimensionless shear rate used in this study (0.1 to 10), the dimensionless timescale is underestimated by a factor of 0.09 to 0.91. 

The Weissenberg number (Wi) \cite{Dealy2010, Blackwell2014, Ewoldt2017}, also named the Mnemosyne number \cite{Jamali2022} when referring to thixotropic responses, quantifies the flow strength and nonlinearity, and here it is defined as
\begin{equation}
    {\rm Wi} = \frac{\dot \gamma}{\dot \gamma_{\rm char}} =  \frac{k_D}{k_A} \dot \gamma =  \dot \gamma ^*.
\end{equation}
The larger the Wi, the larger the structural breakage, $k_D \dot \gamma$, and the less structured the system is~(Eq.~\ref{Eq.thixo_kinetics_dimensional_ss}).

Using the dimensionless time, shear rate, and stress defined above, the dimensionless constitutive equations of the simplest thixotropic kinetic model can be written as:
\begin{align}
     \tau^* &= \xi \dot \gamma ^*, 
     \label{Eq.thixo_stress_dimensionless} \\
    \frac{d\xi}{d t^*} &= 1-\xi-\dot \gamma^* \xi.
    \label{Eq.thixo_structure_dimensionless}
\end{align}

At steady state, the model shows shear-thinning, as shown in Fig.~\ref{Fig.models}\textcolor{blue}{(a)}. At low shear rates, the model approaches a Newtonian limit and the stress shows a slope of one, while at high shear rates, the stress approaches a constant, as the viscosity exhibits an extreme shear-thinning.

\begin{figure}[h!]
	\centering
	\subfigure[]{
		\label{Fig.Thixo_stepshear_structureparameter}
		\includegraphics[width=0.4\textwidth]{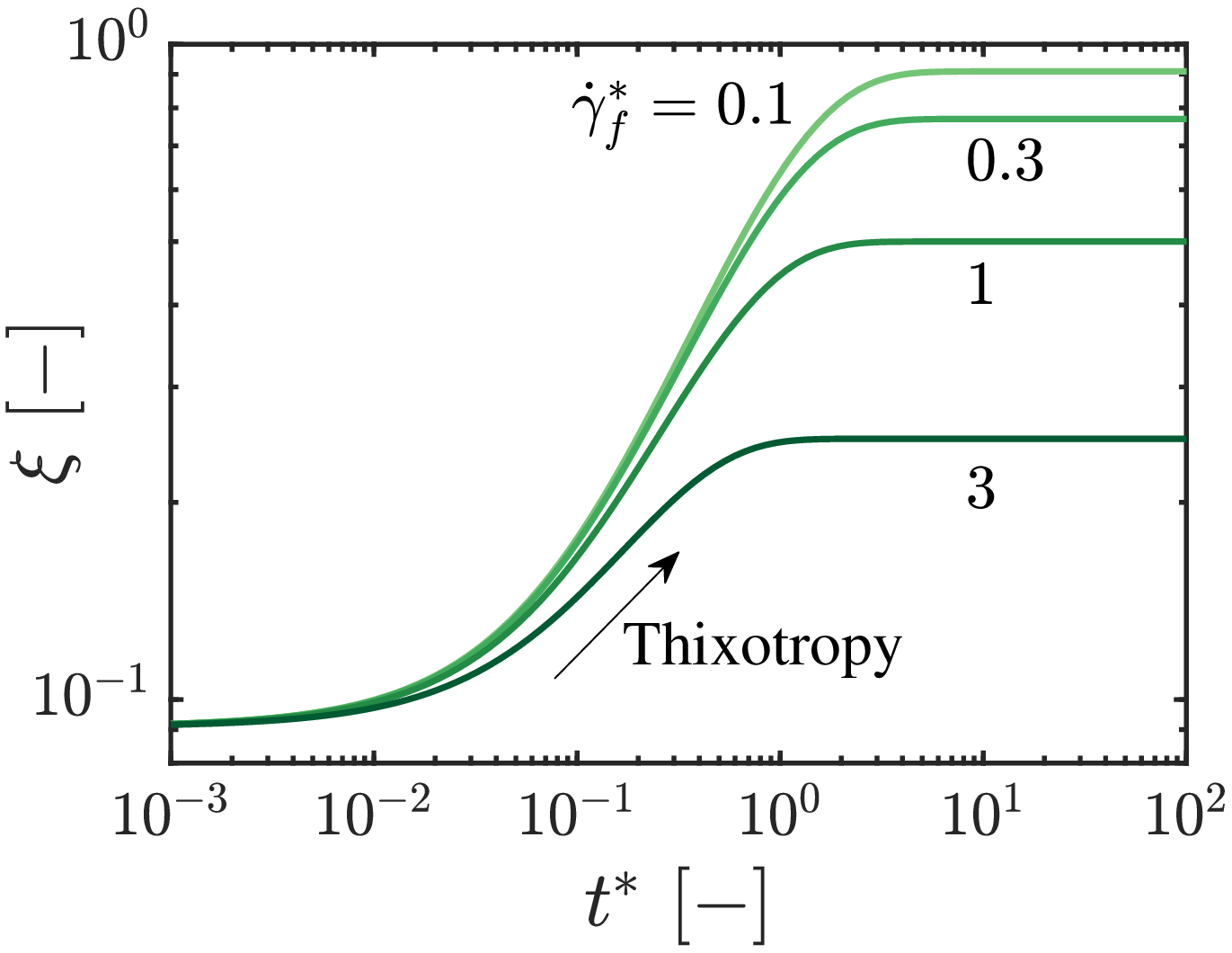}}
	\subfigure[]{
		\label{Fig.Thixo_stepshear_stress}
		\includegraphics[width=0.4\textwidth]{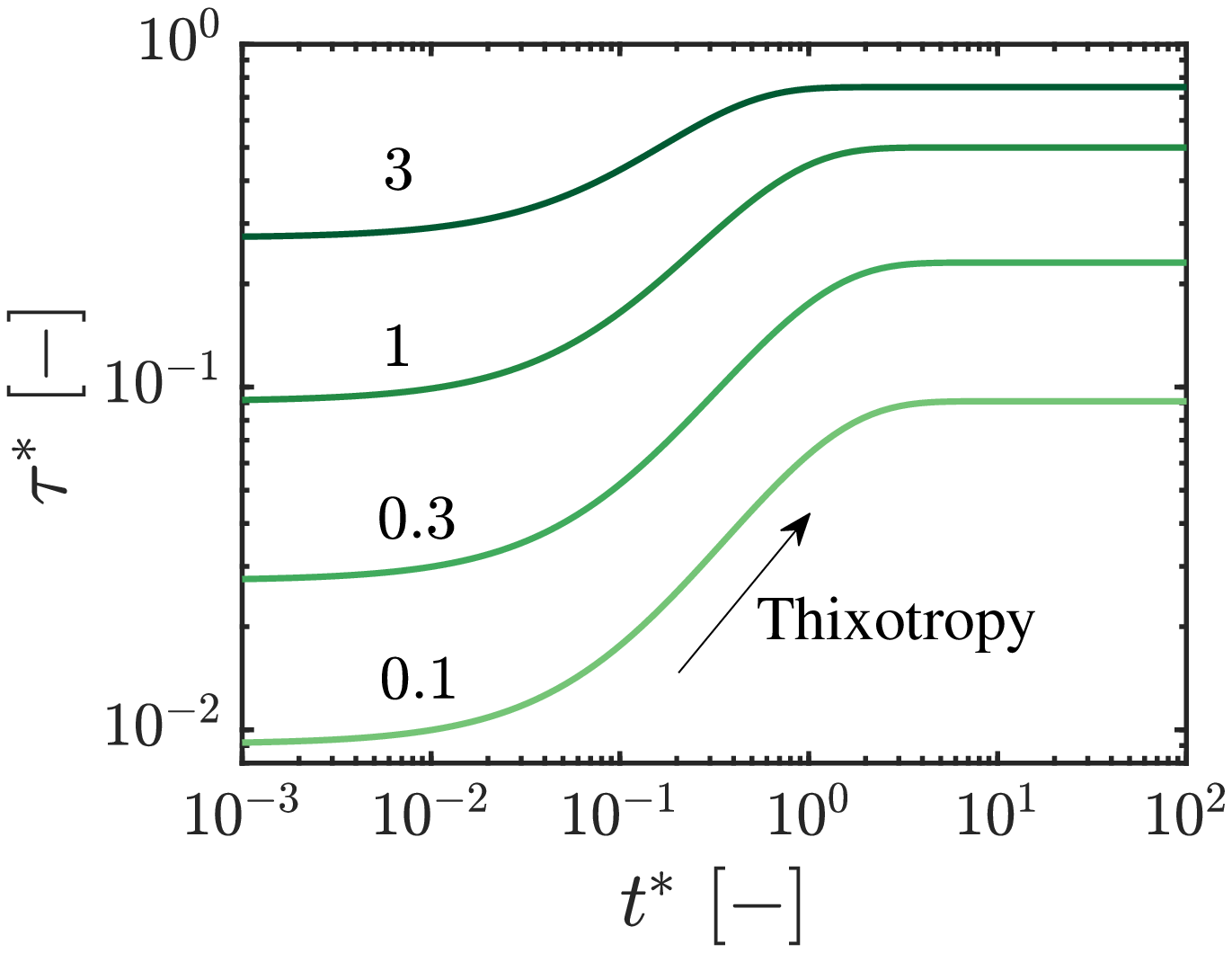}}
	\hspace{0.01\linewidth}
	\caption{Step-down in shear rate tests ($\dot \gamma^*$ \textifsym{\FallingEdge}) for the thixotropic model. (a) Structure parameter and (b) stress each show recovery with time when the shear rate applied is decreased from $\dot \gamma ^*_i = 10$ to $\dot \gamma ^*_f=$ 3, 1, 0.3, and 0.1.}
	\label{Fig.thixo_stepshear}
\end{figure}

When subjected to a step change in shear rate input, where the applied shear rate is decreased from an initial shear rate, $\dot \gamma^*_i$, to a final shear rate, $\dot \gamma^*_f$, at time zero, and with an arbitrary initial condition $\xi \left( t^* = 0\right) = \xi_i$, the structure parameter evolves with time as (solving Eq.~\ref{Eq.thixo_structure_dimensionless})
\begin{equation}
    \xi = \frac{1}{1+\dot \gamma_f^*} + \left( \xi_i -  \frac{1}{1+\dot \gamma_f^*}  \right) {\rm exp} \left( -(1+\dot \gamma_f^*) t^* \right).
\end{equation}
Assuming the system is at steady state under shear rate of $\dot \gamma_i^*$ before stepping down to the final shear rate, we have $\xi_i =  \frac{1}{1+\dot \gamma_i^*}$. The corresponding stress can be calculated using Eq.~\ref{Eq.thixo_stress_dimensionless}. The structure parameter and the stress evolving with time are shown in Fig.~\ref{Fig.thixo_stepshear}, when the applied shear rate is decreased from $\dot \gamma_i^*=10$ to $\dot \gamma^*_f =$ 0.1, 0.3, 1, and 3 respectively. Fig.~\ref{Fig.Thixo_stepshear_structureparameter} shows the response of the thixotropic structure parameter, where from top to bottom, the final shear rate decreases from 3 to 0.1. All transient structure parameters start from the same value at time zero and increase exponentially with time; the smaller the final shear rate, the higher the structure parameter, and the more structured the fluid is. Fig.~\ref{Fig.Thixo_stepshear_stress} shows the corresponding stress evolving with time, where in all cases, stress increases with time, and the smaller the final shear rate, the lower the stress.

\subsection{\label{sec:anti-thixo_model}The simplest anti-thixotropic kinetic model}

Though there is a great number of thixotropic models, studies on anti-thixotropic models are less common. Buitenhuis and P{\"{o}}nitsch \cite{Buitenhuis2003} developed an anti-thixotropic model for polymer solutions, however, the anti-thixotropy only appears at certain shear rates and there is a divergence in viscosity, which complicates the calculation. For simplicity, here, we construct a simple anti-thixotropic kinetic model, using the same idea of the thixotropic model \cite{Larson2015}. It has only one viscous component, as shown in Fig.~\ref{Fig.models}\textcolor{blue}{(b)}, and the viscosity, $\eta$, depends linearly on the anti-thixotropic structure parameter, $\eta = \eta_A \hat \xi$. The anti-thixotropic structure parameter is defined as opposite to the thixotropic structure parameter, 
\begin{equation}
    \hat \xi = 1- \xi,
\end{equation}
and by substituting it into the Moore model, we derive a kinetic equation for anti-thixotropy,
\begin{equation}
    \frac{d \hat \xi}{d t} = - k_0 \hat \xi + k_1 \dot \gamma ( 1- \hat \xi),  
    \label{Eq.antithixo_kinetics}
\end{equation}
where $k_0$ and $k_1$ are constants.

At steady state in Eq.~(\ref{Eq.antithixo_kinetics}), the anti-thixotropic structure parameter depends on the applied shear rate as
\begin{equation}
    \hat \xi_{\rm ss} = \frac{k_1 \dot \gamma}{k_0+ k_1 \dot \gamma} = 
    \frac{ \frac{k_1 \dot \gamma }{k_0} }{1 + \frac{k_1 \dot \gamma }{k_0} }.
\end{equation}
Anti-thixotropic structure $\hat \xi$ is zero at zero shear rate, increases with the applied shear rate and ranges between zero and one. This reveals a dimensionless shear rate
\begin{equation}
    \dot\gamma ^*  = \frac{\dot\gamma}{\dot\gamma_{\rm char} }= \frac{k_1}{k_0} \dot\gamma. 
\end{equation}
The Weissenberg number (Wi) is defined as the dimensionless shear rate $\rm Wi = \dot \gamma ^* = \frac{k_D}{k_A} \dot \gamma $.

For transients, similar to the structure-based thixotropic model, there are two timescales, the zero order timescale, $k_0^{-1}$, which is zero order to shear rate and represents the anti-thixotropic loss of structure at low shear rates. The second timescale is $\left( k_1 \dot \gamma \right)^{-1} $, representing anti-thixotropic structure increase at high shear rates. Similar to the structure-based thixotropic model (Eq.~\ref{Eq.thixo_kinetics}), the dimensionless timescale for anti-thixotropy in hysteresis is defined as
\begin{equation}
    T^* = k_0 T,
\end{equation}
which is an underestimate when shear rate is finite, similar to Eq.~(\ref{Eq.thixo_timescale_definition}). 

The stress of our anti-thixotropic model is
\begin{equation}
    \tau = \eta_A \hat \xi \dot\gamma,
    \label{Eq.anti-thixo_dimensional_stress}
\end{equation}
and the stress is nondimensionalized by a characteristic stress, $\tau_{\rm char} = \eta_A \dot \gamma_{\rm char}$, as 
\begin{equation}
    \tau^* = \frac{\tau}{\tau_{\rm char}} = \frac{k_1}{\eta_A k_0} \tau .
\end{equation}
The steady state stress is a monotonically increasing function of shear rate, and is shear-thickening. In the limit of zero shear rate, the apparent viscosity goes to zero because of Eq.~(\ref{Eq.anti-thixo_dimensional_stress}), which is unrealistic. However, the anti-thixotropic model is still useful for our purposes as it represents the extra stress due to anti-thixotropic dynamics. One could avoid zero apparent viscosity by adding a constant (arbitrarily small) background viscosity in Eq.~(\ref{Eq.anti-thixo_dimensional_stress}). 

With the dimensionless parameters defined above, we can write the dimensionless anti-thixotropic constitutive model as
\begin{align}
    \tau^* &= \hat \xi \dot \gamma ^*, \\
    \frac{d \hat \xi}{d t^*} &= -\hat \xi + \dot \gamma^* (1- \hat \xi)
    \label{Anti_model_dimensionless_structure_kinetic_eq}.
\end{align}
The steady state flow curve of the anti-thixotropic model is shown in Fig.~\ref{Fig.models}\textcolor{blue}{(b)}, where we can see the viscosity increases with the applied shear rate and the model shows a shear-thickening behavior.

\begin{figure}[h!]
	\centering
	\subfigure[]{
		\label{Fig.Antithixo_stepshear_structureparameter}
		\includegraphics[width=0.4\textwidth]{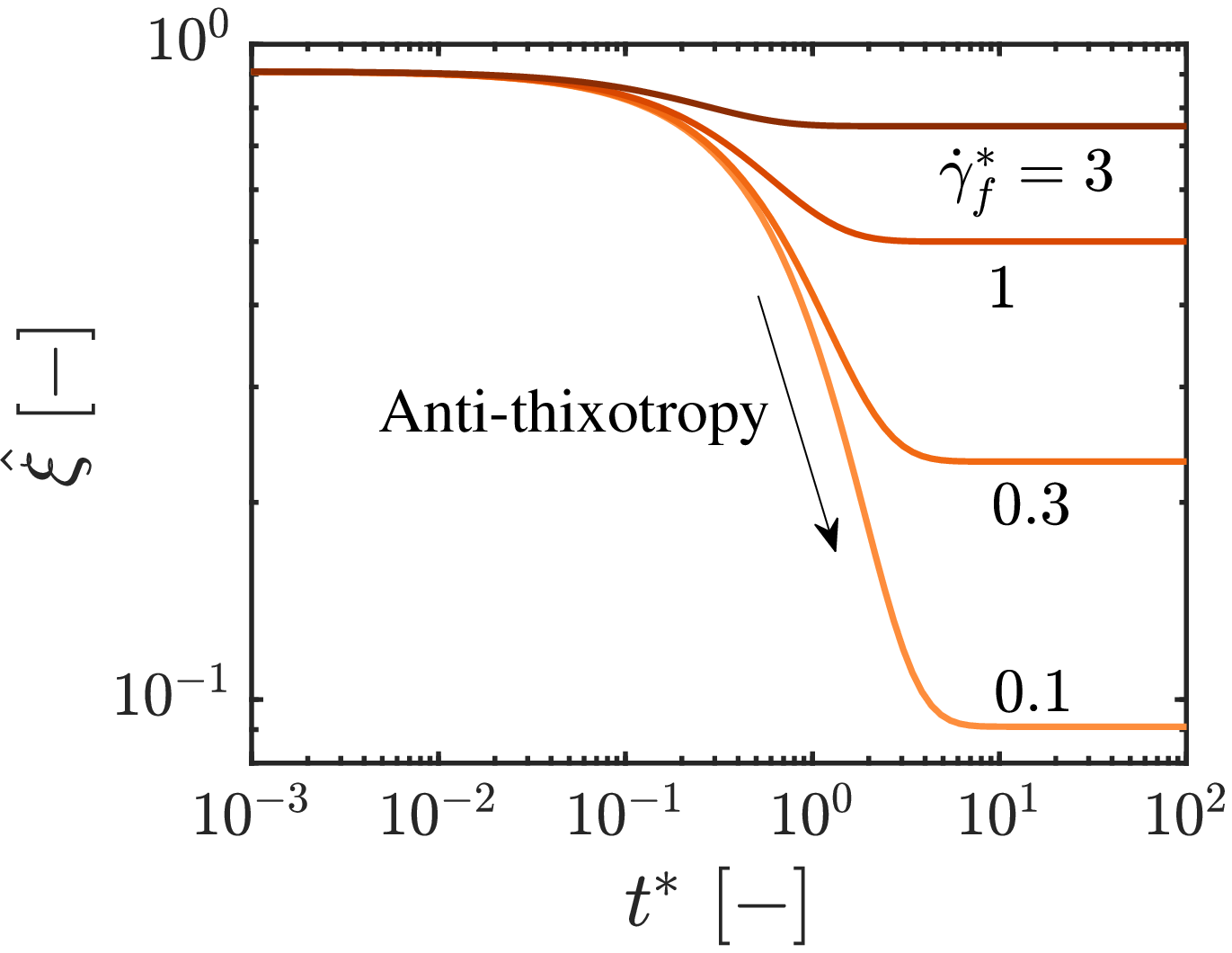}}
	\subfigure[]{
		\label{Fig.Antithixo_stepshear_stress}
		\includegraphics[width=0.4\textwidth]{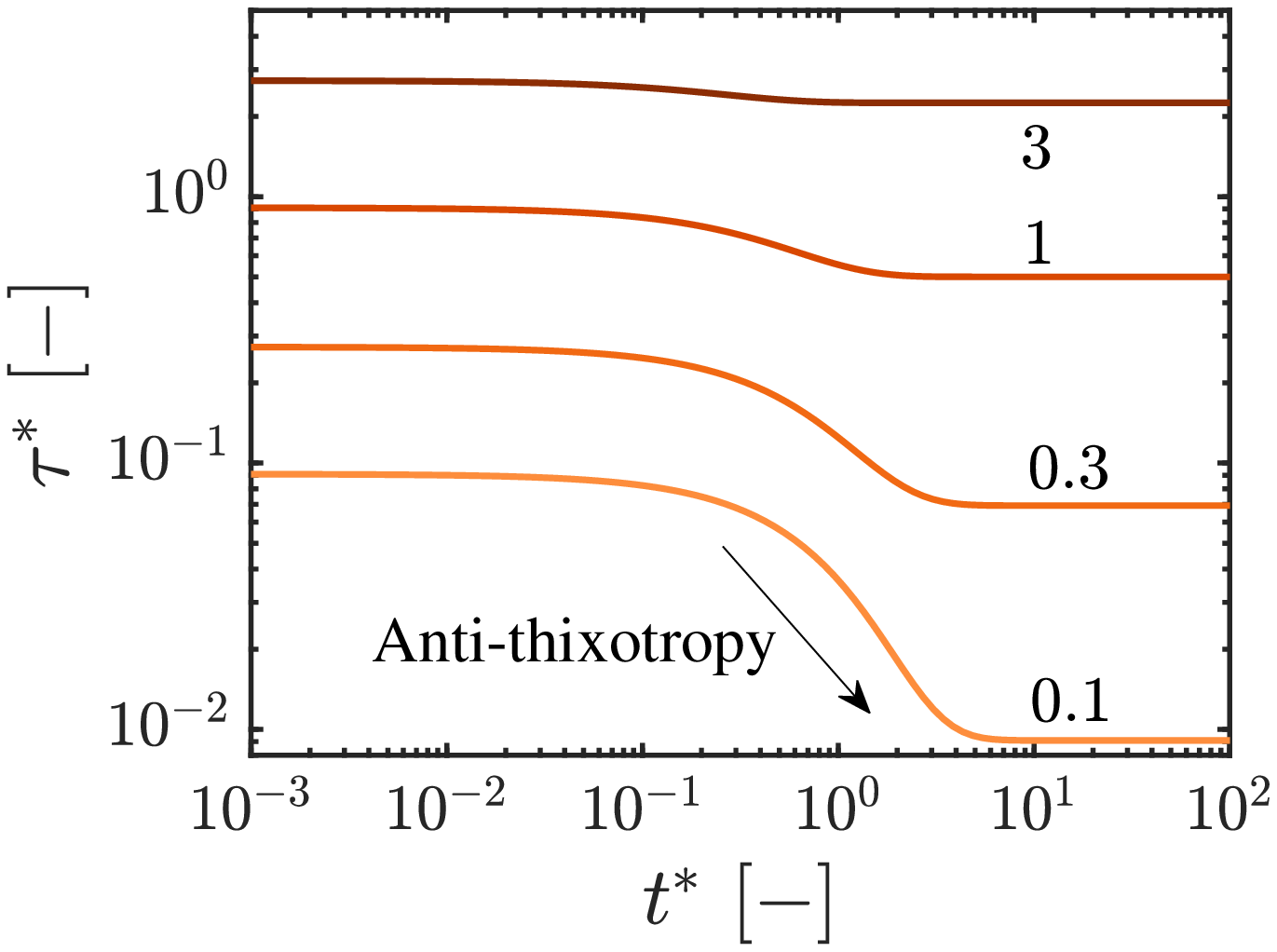}}
	\hspace{0.01\linewidth}
	\caption{Step-down in shear rate tests ($\dot \gamma^*$ \textifsym{\FallingEdge}) for anti-thixotropic model: (a) structure parameter and (b) stress each show decay with time under a step-down in shear rate flow, when the shear rate applied, $\dot \gamma ^*$ is decreased from 10 to 3, 1, 0.3, and 0.1.}
	\label{Fig.antithixo_stepshear}
\end{figure}

When under step changes, where the applied shear rate is decreased from the initial value $\dot \gamma_i^*$ to $\dot \gamma_f^*$ at time zero, and the initial anti-thixotropic structure parameter at time zero is $\hat \xi_i$, the structure parameter evolving with time can be solved from Eq.~(\ref{Anti_model_dimensionless_structure_kinetic_eq}) as
\begin{equation}
    \hat \xi = \frac{\dot \gamma_f^*}{1+\dot \gamma_f^*} + \left( \hat \xi_i -  \frac{\dot \gamma_f^*}{1+\dot \gamma_f^*}  \right) {\rm exp} \left( -(1+\dot \gamma_f^*) t^* \right).
\end{equation}
If the system is at steady state under shear rate of $\dot \gamma_i^*$ at time zero, then $\hat \xi_i =  \frac{\dot \gamma_i^*}{1+\dot \gamma_i^*}$. Fig.~\ref{Fig.antithixo_stepshear} shows the anti-thixotropic structure parameter and shear stress evolving with time. When the applied shear rate is initially $\dot \gamma_i^*=10$, steady state is reached, and the final shear rates, $\dot \gamma_f^*$, are 0.1, 0.3, 1, and 3 respectively. Opposite to the thixotropic model, the anti-thixotropic structure parameter and stress show a decrease with time after the step-down shear rate. The lower the final shear rate, the lower the anti-thixotropic structure parameter, and the lower the stress.

\subsection{\label{sec:Giesekus_model}The nonlinear viscoelastic Giesekus model}

There are many nonlinear viscoelastic models available in literature \cite{DPL1987}, and several have been used to calculate hysteresis \cite{Rubio2008, Marsh1968, Bird1968, Sharma2022}. In this study, the Giesekus model is used to study nonlinear viscoelasticity in hysteresis, because it shows shear-thinning, which makes its nonlinear time dependence harder to be differentiated from thixotropy. The stress tensor of the Giesekus model \cite{DPL1987} contains two parts, a viscoelastic stress from a polymer contribution, $\underline{\underline{\tau_p}}$, and a Newtonian stress from a solvent contribution, $\underline{\underline{\tau_s}}$. Consistent with the schematics of the Giesekus model in Fig.~\ref{Fig.models}\textcolor{blue}{(c)}, stresses are additive,     
\begin{equation}
   \underline{\underline{\tau}} = \underline{\underline{\tau_s}} + \underline{\underline{\tau_p}}.  
   \label{Eq.Giesekus_dimensional_1}
\end{equation}
The Newtonian stress is proportional to the shear rate tensor,
\begin{equation}
    \underline{\underline{\tau_s}} = \eta_s \underline{\underline{\dot \gamma}},
    \label{Eq.Giesekus_dimensional_2}
\end{equation}
where $\eta_s$ is the viscosity of the solvent. The viscoelastic stress is governed by the constitutive equation,
\begin{equation}
    \underline{\underline{\tau_p}} +  \lambda_1\underline{\underline{\tau_p}}_{(1)} +  \alpha \frac{\lambda_1}{\eta_p} \left\{ \underline{\underline{\tau_p}} \cdot \underline{\underline{\tau_p}} \right\} = \eta_p \underline{\underline{\dot \gamma}},
    \label{Eq.Giesekus_dimensional_3}
\end{equation}
where $\alpha$ is the dimensionless "mobility factor" that can be associated with anisotropic Brownian motion and/or anisotropic hydrodynamic drag on the constituent polymer molecules, which ranges from 0 to 0.5, $\eta_p$ is the polymer viscosity at zero shear rate, and $\lambda_1$ is the relaxation time. $\underline{\underline{\tau_p}}_{(1)}$ is the first order upper-convected derivative of $\underline{\underline{\tau_p}}$, which is defined as
\begin{equation}
    \underline{\underline{\tau_p}}_{(1)} = \frac{D}{Dt} \underline{\underline{\tau_p}} - \left( \underline{\nabla}\,\underline{v} \right)^T \cdot \underline{\underline{\tau_p}}  - \underline{\underline{\tau_p}} \cdot \left( \underline{\nabla}\,\underline{v} \right).
\end{equation}

Equations (\ref{Eq.Giesekus_dimensional_1}) - (\ref{Eq.Giesekus_dimensional_3}) can be written in a single equation for the total stress. In this study, we only work with shear rate controlled hysteresis, and the Giesekus model in stress summation form (Eq. \ref{Eq.Giesekus_dimensional_1} - \ref{Eq.Giesekus_dimensional_3}) is more convenient for calculation. Therefore, we only consider these equations in their dimensionless forms.  

The stress summation form of the Giesekus model has three dimensional parameters, $\eta_s$, $\eta_p$ and $\lambda_1$, (another parameter $\alpha$ is dimensionless). There are also three parameters for the hysteresis calculation, $t$, $\underline{\underline{\dot \gamma}}$, and $\underline{\underline{\tau}}$. This yields a total of six parameters with two linearly independent dimensions (all have units of [Pa] or [s] or linear combinations of both). Therefore, by the Buckingham pi theorem \cite{Buckingham1914}, the model can be non-dimensionalized using four dimensionless groups, which are chosen as: 
\begin{align}
    t^*  &= \frac{t}{\lambda_1}, \\
    \underline{\underline{\dot \gamma^*}} &= \lambda_1  \underline{\underline{\dot \gamma}}, \\
    \underline{\underline{\tau^*}} &= \frac{\lambda_1}{\eta_p} \underline{\underline{\tau}}, \\
    \eta_s^* &= \frac{\eta_s}{\eta_p}. 
\end{align}
In linear viscoelasticity, only one timescale governs the stress recovery and relaxation, therefore, we define the dimensionless timescale by the viscoelastic relaxation time, $\lambda_{\rm VE} = \lambda_1$
\begin{equation}
    T^* = \frac{\rm ramping~time~per~decade}{\rm viscoelastic~relaxation~time} = \frac{T}{\lambda_1}.
\end{equation}
The viscoelastic Deborah number is therefore ${\rm De} = 1/T^*$. The viscoelastic Weissenberg number is defined as ${\rm Wi} =\lambda_1 \dot \gamma$, consistent with $\dot \gamma_{\rm char} = 1/ \lambda$, being the critical shear rate to observe nonlinearity in this model \cite{DPL1987}. 

Under homogeneous simple shear, the dimensionless applied shear rate tensor is
\begin{equation}
    \underline{\underline{\dot \gamma^*}} = \left[\begin{array}{ccc}
0 & \dot \gamma^* & 0 \\
\dot \gamma^* &  0 &0 \\
 0 & 0 &  0 
\end{array}\right],
\end{equation}
and the polymer contribution stress tensor, $\underline{\underline{\tau_p^*}}$, can be written as
\begin{equation}
    \underline{\underline{\tau_p^*}} =\left[\begin{array}{ccc}
\tau_{p,xx}^* & \tau_{p,yx}^* & 0 \\
 \tau_{p,yx}^* &  \tau_{p,yy}^* &0 \\
 0 & 0 &  \tau_{p,zz}^* 
\end{array}\right].
\end{equation}
Substituting into Eq.~(\ref{Eq.Giesekus_dimensional_3}) to solve for the stress,
\begin{align}
    \tau_{p,xx}^* + \frac{\partial}{\partial t} \tau_{p,xx}^* - 2 \dot \gamma^* \tau_{p,yx}^* + \alpha ( \tau_{p,xx}^{*2} + \tau_{p,yx}^{*2}) &= 0, 
    \label{Eq.Giesekus_dimensionless_1} \\
    \tau_{p,yy}^* + \frac{\partial}{\partial t} \tau_{p,yy}^*  + \alpha ( \tau_{p,yy}^{*2} + \tau_{p,yx}^{*2} ) &= 0, \\
    \tau_{p,zz}^* + \frac{\partial}{\partial t} \tau_{p,zz}^*  + \alpha \tau_{p,zz}^{*2}  &= 0, \\
    \tau_{p,yx}^* + \frac{\partial}{\partial t} \tau_{p,yx}^* - \dot \gamma^* \tau_{p,yy}^*  + \alpha ( \tau_{p,xx}^{*} \tau_{p,yx}^{*} + \tau_{p,yx}^{*} \tau_{p,yy}^{*} ) &= \dot \gamma^*.
\end{align}
The solvent contribution and the total stress are
\begin{align}
    \underline{\underline{\tau_s^*}} &= \eta_s^* \underline{\underline{\dot \gamma^*}}, \\ 
    \underline{\underline{\tau^*}} &= \underline{\underline{\tau_s^*}} +\underline{\underline{\tau_p^*}}.
    \label{Eq.Giesekus_dimensionless_6}
\end{align}
The shear stress, $\tau_{yx}^*$, denoted as $\tau^*$, therefore can be solved by Eq.~(\ref{Eq.Giesekus_dimensionless_1}) - (\ref{Eq.Giesekus_dimensionless_6}), under given model parameters and applied shear rate and time conditions. In this study, we use $\eta_s^* = 0.1, ~\alpha =0.3$.    

At steady state, the dimensionless shear stress and viscosity as a function of shear rate are calculated and plotted, as shown in Fig.~\ref{Fig.models}\textcolor{blue}{(c)}. The model shows shear-thinning at high shear rates, and at low rates, there exists a Newtonian plateau. The Giesekus model also gives a non-zero first normal stress difference, which is not plotted here. 

\begin{figure}[h!]
	\centering
		\includegraphics[width=0.4\textwidth]{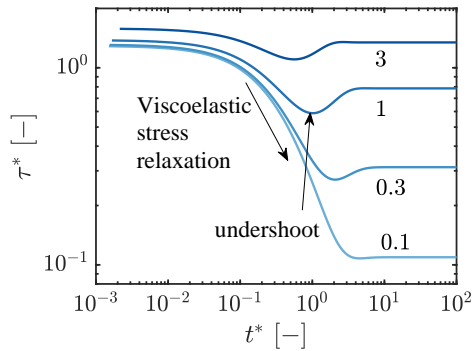}
	\hspace{0.01\linewidth}
	\caption{The nonlinear viscoelastic Giesekus model shows a stress decrease with time under a step-down in shear rate flow ($\dot \gamma^*$ \textifsym{\FallingEdge}), when the shear rate applied, $\dot \gamma ^*$ is decreased from 10 to 3, 1, 0.3, and 0.1. For large shear rates, a stress undershoot is observed during stress decrease.}
	\label{Fig.Giesekus_stepdown}
\end{figure}

Under a step-down in shear rate test, where before time zero, all stress components reach the steady state at the initial shear rate applied, $\dot \gamma_i^*$, the applied shear is decreased a lower value, $\dot \gamma_f^*$, at time zero, and the shear stress, $\tau^*$, evolving with time can be solved numerically (we use the ode45 function by MATLAB \cite{matlab}). Fig.~\ref{Fig.Giesekus_stepdown} shows the shear stress decay function of the Giesekus model, when $\dot \gamma^*$ is decreased from 10 to 3, 1, 0.3, and 0.1. The stress shows a decrease with time under step down in shear rate and reaches the steady state. When the final shear rate is high (large Wi number), there exists a stress undershoot, while at lower shear rate (small Wi number, 0.1), the undershoot becomes negligible. The stress undershoot is a nonlinear response in the shear stress decay function, which has been observed and studied in many nonlinear viscoelastic materials and models \cite{Saengow2019}. It should be noted that under step-down in shear rate test, both anti-thixotropy and viscoelasticity show a stress decrease with time, making them indistinguishable from a single test. Furthermore, the stress undershoot observed in nonlinear viscoelastic systems makes it difficult to be distinguished from real thixotropic materials.

\section{\label{sec:results_model}Fingerprints of Hysteresis}
In this section, we show hysteresis loops of the three models (Sec.~\ref{sec:models}) using the method discussed in Sec.~\ref{sec:hysteresis} and we compare the distinguishing features of them.

\subsection{\label{sec:results_model_hysteresis}The hysteresis loops}

\begin{figure}[h!]
	\centering
	\includegraphics[width=0.8\textwidth]{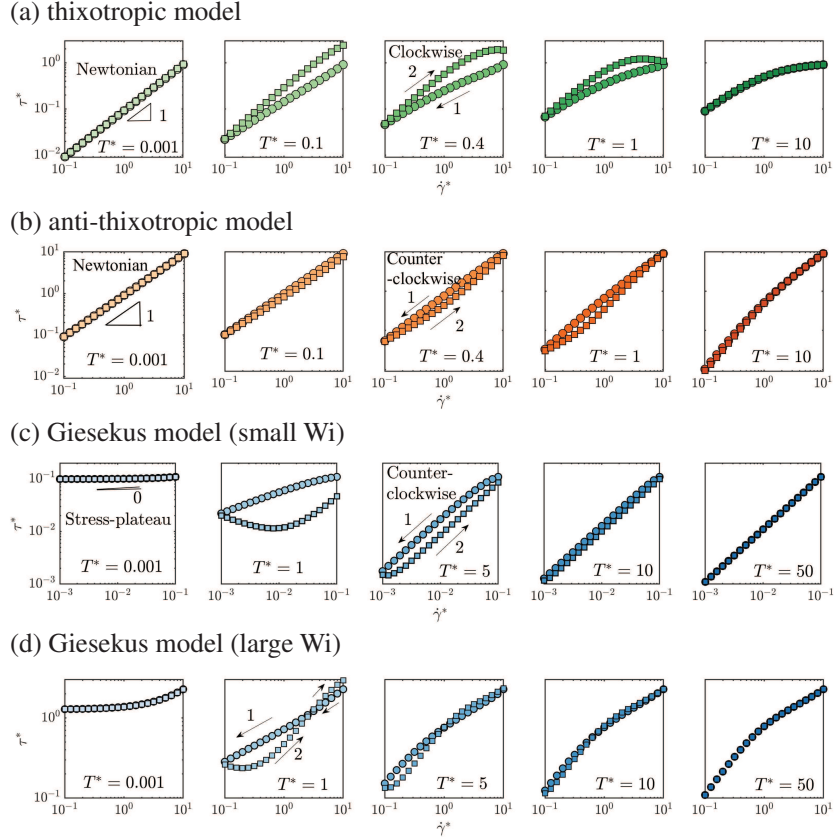}
	\hspace{0.01\linewidth}
	\caption{Hysteresis loops (ramping high-low-high in $\dot \gamma^*$) for (a) the thixotropic model, (b) the anti-thixotropic model, (c) the Giesekus model at small Wi ($\dot \gamma^*_{\rm max} = 0.1$), and (d) the Giesekus model at large Wi ($\dot \gamma^*_{\rm max} = 10$). From left to right, the dimensionless ramping time per decade, $T^*$, increases, and the ramping rate decreases.}
	\label{Fig.comparison_model}
\end{figure}

Fig.~\ref{Fig.comparison_model}, from top to bottom, shows the hysteresis loops for the thixotropic, anti-thixotropic, Giesekus models at small Wi, and Giesekus model at large Wi. From left to right, the ramping time per decade, $T^*$, is increasing as the symbol color gets darker. The downward ramping is plotted as circles, also labeled "1" with arrow pointing down in some plots, and the upward ramping is plotted as squares, labeled "2" with arrow pointing up.

Fig.~\ref{Fig.comparison_model}\textcolor{blue}{(a)} shows the hysteresis loops for the thixotropic model, and from left to right, $T^*$ increases from 0.001 to 10 ($T^* = k_A T$ for this model). Hysteresis loops are only observable at intermediate ramping times and the area of the loop first increases then decreases with $T^*$. At fast ramping, $T^* = 0.001$, the structure parameter, $\xi$, does not have enough time to respond to the change in shear rate, so it remains a constant during the ramping. The small change of structure parameter during hysteresis at $T^*=0.001$ is shown in Fig.~\textcolor{blue}{S5(a)}. As a result, the viscosity remains nearly constant during the ramping, therefore the downward and upward ramping stresses essentially overlap, and the model shows Newtonian behavior. As $T^*$ increases, the area of the hysteresis loop increases, and significant hysteresis appears. The direction of the loop, when observable, is clockwise, that is, the downward ramping curve is lower than the upward ramping. This is because during the downward ramping, at each step, the structure of the thixotropic fluid is building up, and the structure parameter and viscosity are increasing. Therefore, the flow curve at upward ramping (labeled as "2") is higher, resulting in a clockwise loop. The structure parameter during hysteresis ramping at $T^*=0.4$ can be seen at Fig.~\textcolor{blue}{S5(b)}. As $T^*$ keeps increasing, the area of the hysteresis loop begins to decrease and eventually, no hysteresis can be observed, as the ramping time increases to $T^*=10$. For the long enough ramping time, at each step, the structure parameter of the fluid has enough time to reach steady state, as shown in Fig.~\textcolor{blue}{S5(c)}. Therefore, the downward and upward ramping curves overlap, and reduce to the steady state flow curve, as shown in Fig.~\ref{Fig.models}\textcolor{blue}{(a)}, which yields Newtonian behavior at low shear rates and extreme shear-thinning at high rates. 

The hysteresis loops for the anti-thixotropic model at different $T^*= k_0 T $ ranging from 0.001 to 10 are shown in Fig.~\ref{Fig.comparison_model}\textcolor{blue}{(b)}. Similar to that of the thixotropic model, from left to right, $T^*$ increases as the symbols get darker. The hysteresis is only observable at intermediate ramping times, and the area first increases then decreases with $T^*$. At very quick ramping, $T^*=0.001$, the anti-thixotropic structure parameter, $\hat \xi$, remains nearly constant, therefore the downward and upward stress curves overlap and show Newtonian behavior. At an intermediate ramping time, $T^*=0.4$, at each step during the downward ramping, $\hat \xi$ and $\tau^*$ decrease with time. Therefore, the upward ramping has a lower stress compared with the downward ramping, giving a counter-clockwise hysteresis loop. As $T^*$ gets larger, the steady state can be achieved at each step during the ramping, therefore, the downward and upward ramping curves overlap and reduce to the steady state flow curve shown in Fig.~\ref{Fig.models}\textcolor{blue}{(b)}.

The hysteresis loops for the nonlinear viscoelastic Giesekus model show a strong shear rate dependence. This is different from both thixotropic and anti-thixotropic models, where the qualitative characteristic features of hysteresis loops do not depend on the range of shear rate applied because of the continuous change of structure parameter, i.e. no self-intersection is induced so long as the ramping starts from an equilibrium condition (after enough time of pre-shear at the maximum shear rate). However, for the nonlinear viscoelastic Giesekus model, the features are significantly different in linear (small shear rate, small Wi) and nonlinear (large shear rate, large Wi) regimes. This can be expected from the stress response in a step-down shear rate test, which shows different trends in linear and nonlinear regimes. At low shear rates (low Wi, linear regime), stress decreases monotonically, while at high shear rates (high Wi, nonlinear regime), significant stress undershoot can be observed. Therefore, hysteresis loops for the Giesekus model need to be considered in both linear and nonlinear regimes.

Fig.~\ref{Fig.comparison_model}\textcolor{blue}{(c)} shows the hysteresis loops for the Giesekus model in the linear regime (small Wi), where the shear rate is ramped between $\dot \gamma^*_{\rm max} = 0.1$ to $\dot \gamma^*_{\rm min} = 0.001$. As $T^* = T/ \lambda$ increases from 0.001 to 50 from left to right, the area of the hysteresis loop increases, then decreases. No hysteresis is observed at very quick ramping, and the stress remains a constant for all shear rates, showing a peculiar stress plateau at $T^*=0.001$. This is because at very quick ramping, the elastic stress does not have enough time to relax and respond to the decreasing shear rate. With the increase in $T^*$, hysteresis area increases, and the direction of the hysteresis loop, when observable, is counter-clockwise. As $T^*$ keeps increasing, the stress has enough time to relax to steady state at each step, the area becomes negligible, the hysteresis reduces to the steady state flow curve shown in Fig.~\ref{Fig.models}\textcolor{blue}{(c)}, showing a Newtonian behavior in the shear rates tested (low Wi regime).      

The shapes of the hysteresis loops for the Giesekus model become more complicated at large shear rates (nonlinear, high Wi regime), as shown in Fig.~\ref{Fig.comparison_model}\textcolor{blue}{(d)}, the hysteresis loops for shear rate ramped from $\dot \gamma^*_{\rm max} = 10$ to $\dot \gamma^*_{\rm min} = 0.1$ at $T^*$ ranging from 0.001 to 50. Similar to the linear regime, no hysteresis loop can be observed at very quick (short $T^*$) and slow (long $T^*$) rampings. At $T^*=0.001$, the flow curve again shows a peculiar pseudo-plastic behavior with a stress plateau at low shear rates. Although this may appear as shear-thinning pseudo plastic behavior from the view of $\tau^* (\dot \gamma ^*)$, the stress plateau is actually elastic and constant due to insufficient time to allow relaxation. The rate dependence at high $\dot \gamma^*$ is due to the solvent contribution to the shear stress, which gives a Newtonian behavior, in superposition to the polymer contribution. At intermediate ramping times, hysteresis is observed, however, self-intersection also appears, which changes the direction of the hysteresis loop from counter-clockwise at low shear rates to clockwise at high shear rates. The appearance of intersection is a result of stress undershoot at step change tests, similar to Fig.~\ref{Fig.Giesekus_stepdown}, which only appears at high shear rates (large Wi, nonlinear regime). As the time $T^*$ keeps increasing, steady state is achieved and the stress curves overlap and recover the steady state flow curve, Fig.~\ref{Fig.models}\textcolor{blue}{(c)}. 

Comparing the fingerprints of hysteresis loops for the three models, some distinguishing features can be identified. The first is a binary feature of clockwise versus counter-clockwise loops: clockwise for thixotropy, but counter-clockwise for viscoelasticity (in low Wi) and anti-thixotropy. A second feature is therefore required to distinguish the latter two dynamics, and this is achieved at high ramping rates ($T^* \ll 1$), where hysteresis is removed and both thixotropic and anti-thixotropic models show nearly Newtonian behavior with minimal shear-thinning/thickening due to constant structure, whereas the viscoelastic model shows a stress plateau at low shear rates. We interpret $T^* \ll 1$ as either a high viscoelastic Deborah number for which elastic stress does not have time to relax, or a high thixotropic Deborah number for which the thixotropic structure does not have time to change. The distinguishing features are observed independent of the model details. Apart from these two distinguishing features, those most basic thixotropic and anti-thixotropic models do not have self-intersection, while the nonlinear viscoelastic Giesekus model demonstrates a shear rate dependence: at low shear rates, (low Wi), there is no self-intersection, while at high shear rates, (high Wi), there is one self-intersection that changes the direction of the hysteresis loop in the high shear rate regime.  

\subsection{\label{sec:results_model_rampingdown}The hysteresis downward ramping}

\begin{figure}[h!]
	\centering
	\subfigure[]{
		\label{Fig.thixotropy_hysteresis_rampingdown_stress_ss}
		\includegraphics[width=0.3\textwidth]{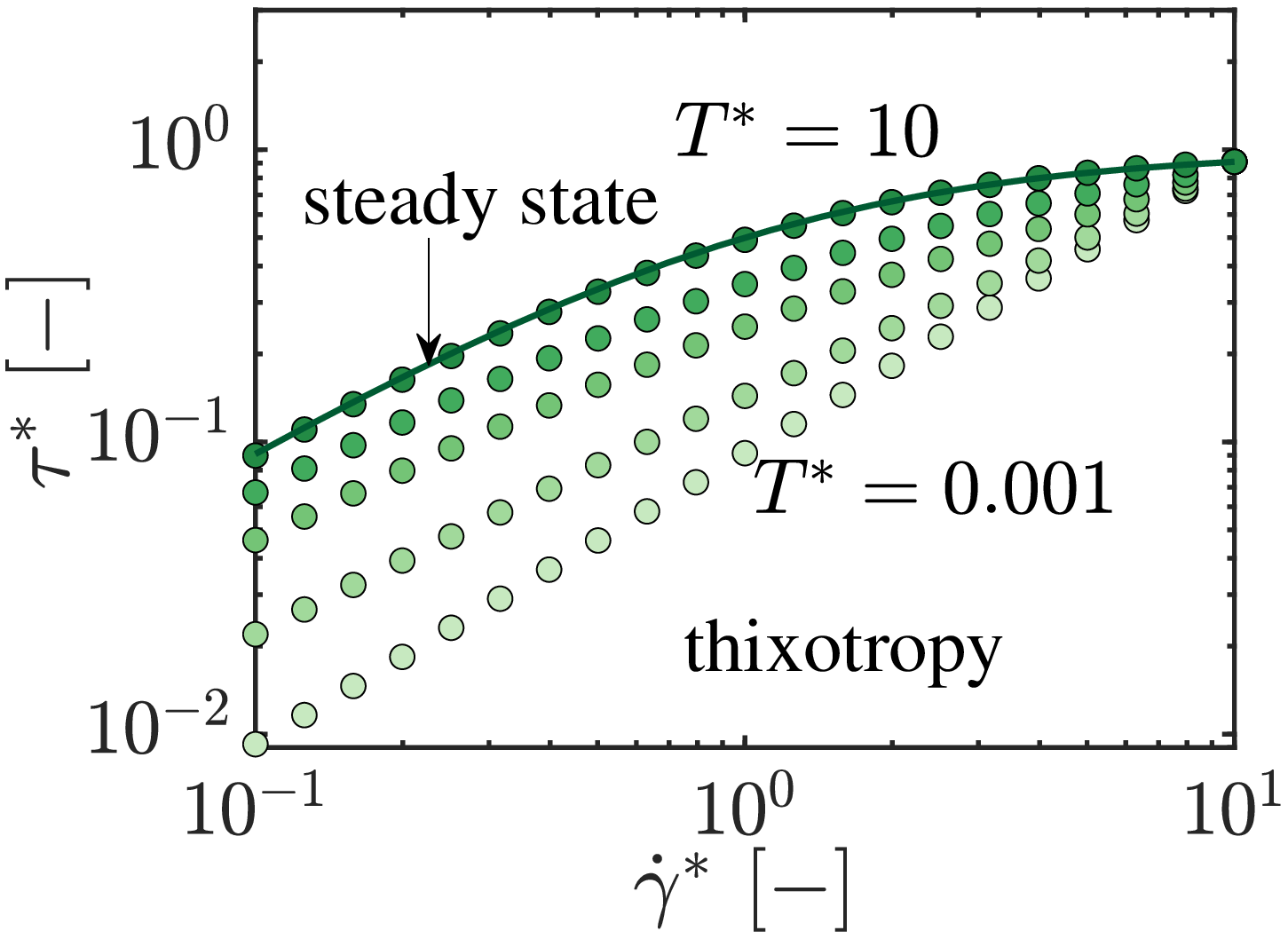}}
	\subfigure[]{
		\label{Fig.antithixotropy_hysteresis_rampingdown_stress_ss}
		\includegraphics[width=0.3\textwidth]{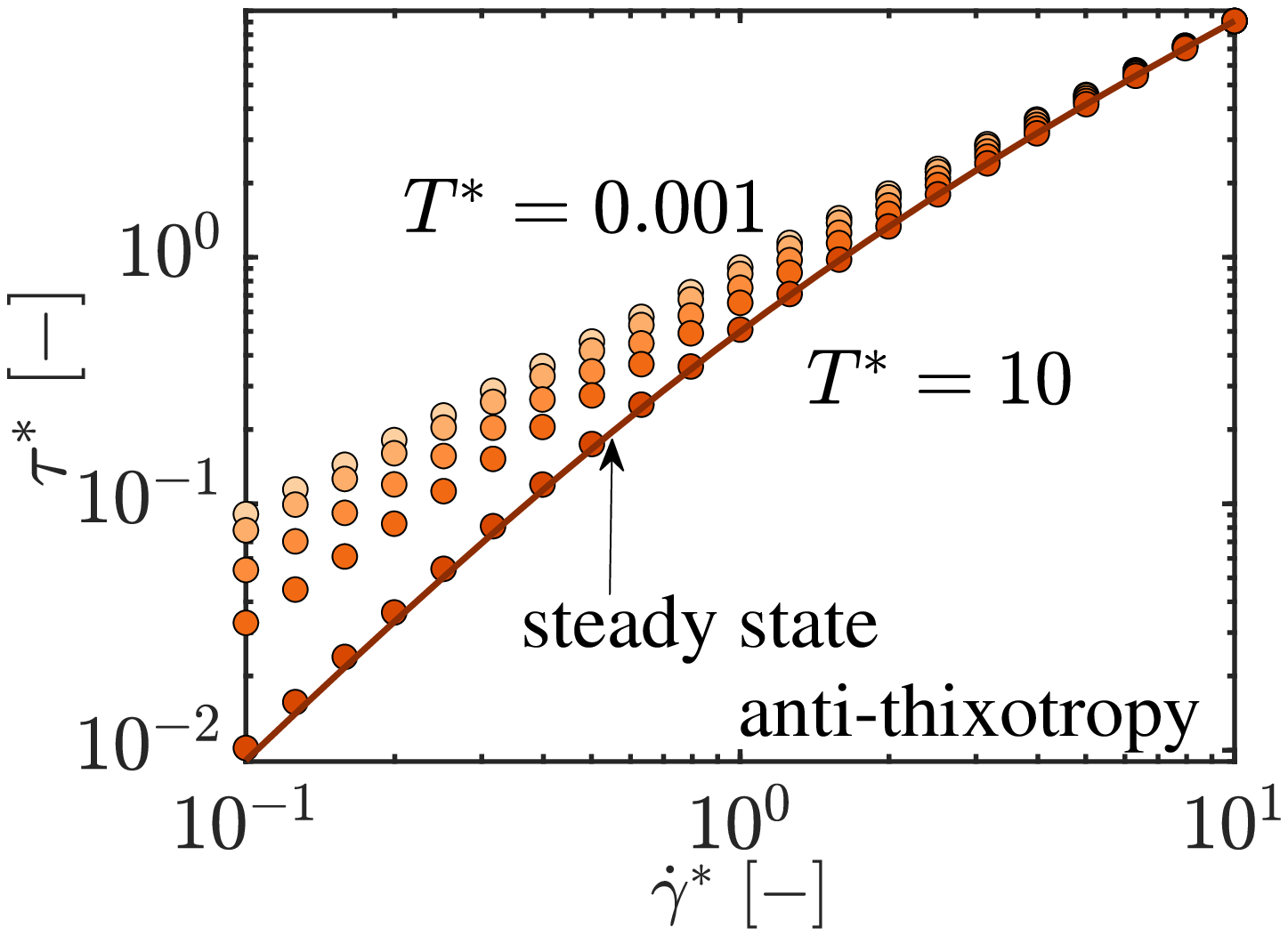}}
	\hspace{0.01\linewidth}
		\subfigure[]{
		\label{Fig.Giesekus_hysteresis_rampingdown_stress_ss}
		\includegraphics[width=0.3\textwidth]{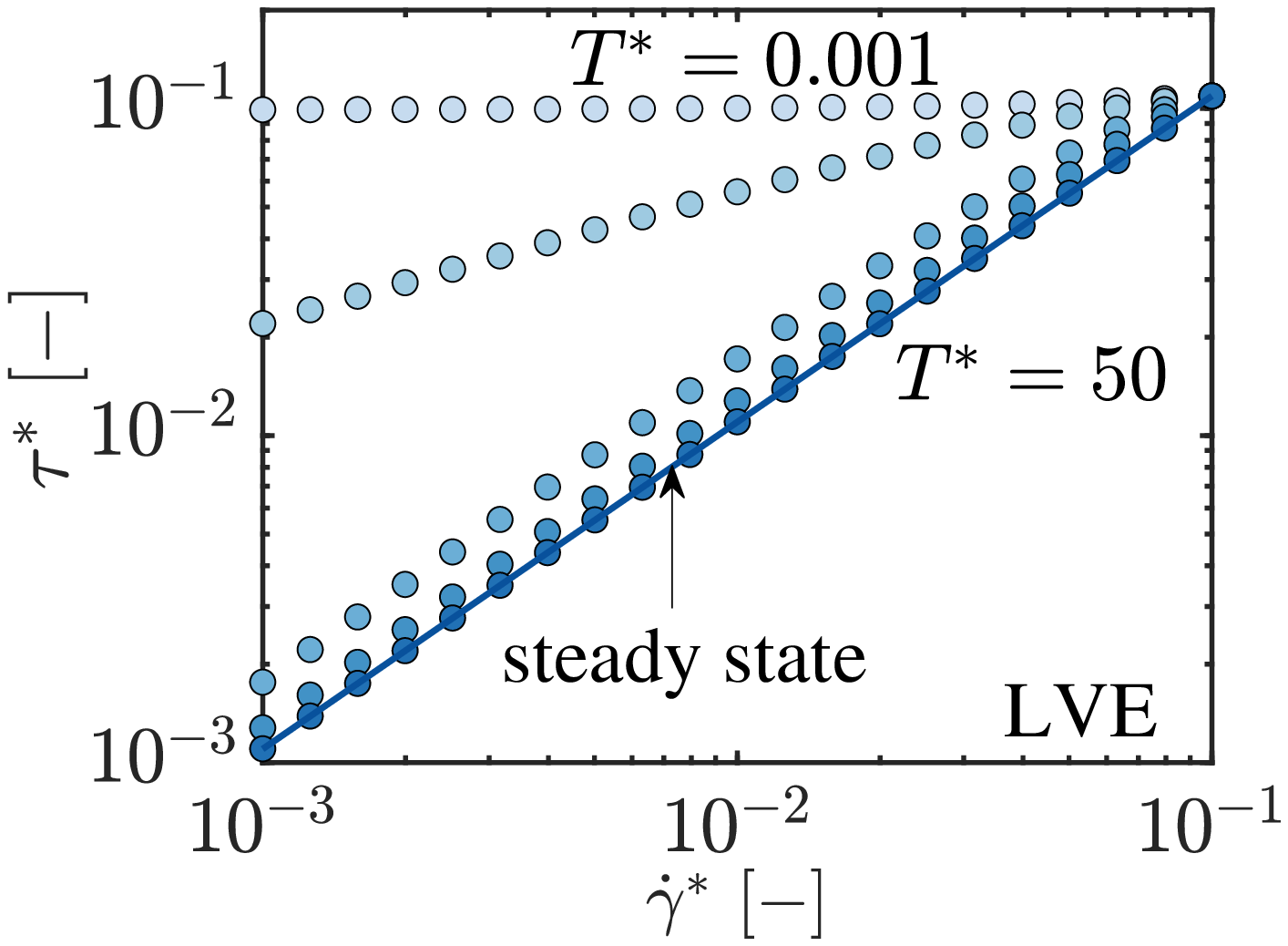}}
	\hspace{0.01\linewidth}
	\caption{The change in the stress with different $T^*$ in ramping down sweep, transient stress versus shear rate for (a) the thixotropic model, (b) the anti-thixotropic model, and (c) the Giesekus model at linear regime. The solid line shows the steady state flow curve for each model; as $T^*$ increases, the ramping down curve is approaching the steady state.}
	\label{Fig.ramping_down}
\end{figure}

Another distinguishing feature can be observed when one looks at the downward rampings. Fig.~\ref{Fig.ramping_down} shows the flow curves at downward ramping at different ramping rates for the three models. Consistent with the legends used in Fig.~\ref{Fig.comparison_model}, the darker the symbol, the longer the ramping time per decade ($T^*$ increasing). The solid lines show the steady state flow curves. From Fig.~\ref{Fig.ramping_down}, for the thixotropic model, all the hysteresis ramping down curves are below the steady state flow curve, and the longer the $T^*$, the higher the stress. While for the anti-thixotropic and viscoelastic Giesekus model in linear regime, hysteresis stresses are higher than the steady state stresses, and with increasing $T^*$, the stress decreases and approaches the steady state. This is another characteristic feature that can distinguish thixotropy from the other two dynamics. 

\subsection{\label{sec:results_model_area}The hysteresis area}

Hysteresis loops can also be used to quantify the intrinsic timescale of the dynamics. As observed in Sec.~\ref{sec:results_model_hysteresis}, for all three models, the area of hysteresis loops changes with the ramping time per decade, $T^*$, and it maximizes at an intermediate $T^*$ and reduces to zero at extreme small and large $T^*$. It has been found that the area of the thixotropic hysteresis loop is maximized when the thixotropic timescale is comparable to the ramping time \cite{Divoux2013, Radhakrishnan2017, Jamali2019, Puisto2015}. In this section, we analyze the dimensionless hysteresis area, $A_\tau^*$, as a function of $T^*$ quantitatively. This reveals a characteristic timescale of the models and allows the comparison between different models. The area of the hysteresis loop is calculated \cite{Divoux2013,Jamali2022} as
\begin{equation}
    A_\tau^* = \int_{\dot\gamma^*_{\rm min}}^{\dot \gamma^*_{\rm max}} \left| \Delta \tau^* \left( \dot \gamma^* \right) \right| \,d \left(  {\rm log} \dot \gamma^*  \right),
\end{equation}
where $ \Delta \tau^* \left( \dot \gamma^* \right) = \tau^*_{\rm up}\left( \dot \gamma^* \right) - \tau^*_{\rm down}\left( \dot \gamma^* \right) $. 
This definition was suggested by Divoux and coworkers to quantify the amplitude of the hysteresis phenomenon \cite{Divoux2013}.

\begin{figure}[h!]
	\centering
	\subfigure[]{
		\label{Fig.thixotropy_hysteresis_area}
		\includegraphics[width=0.45\textwidth]{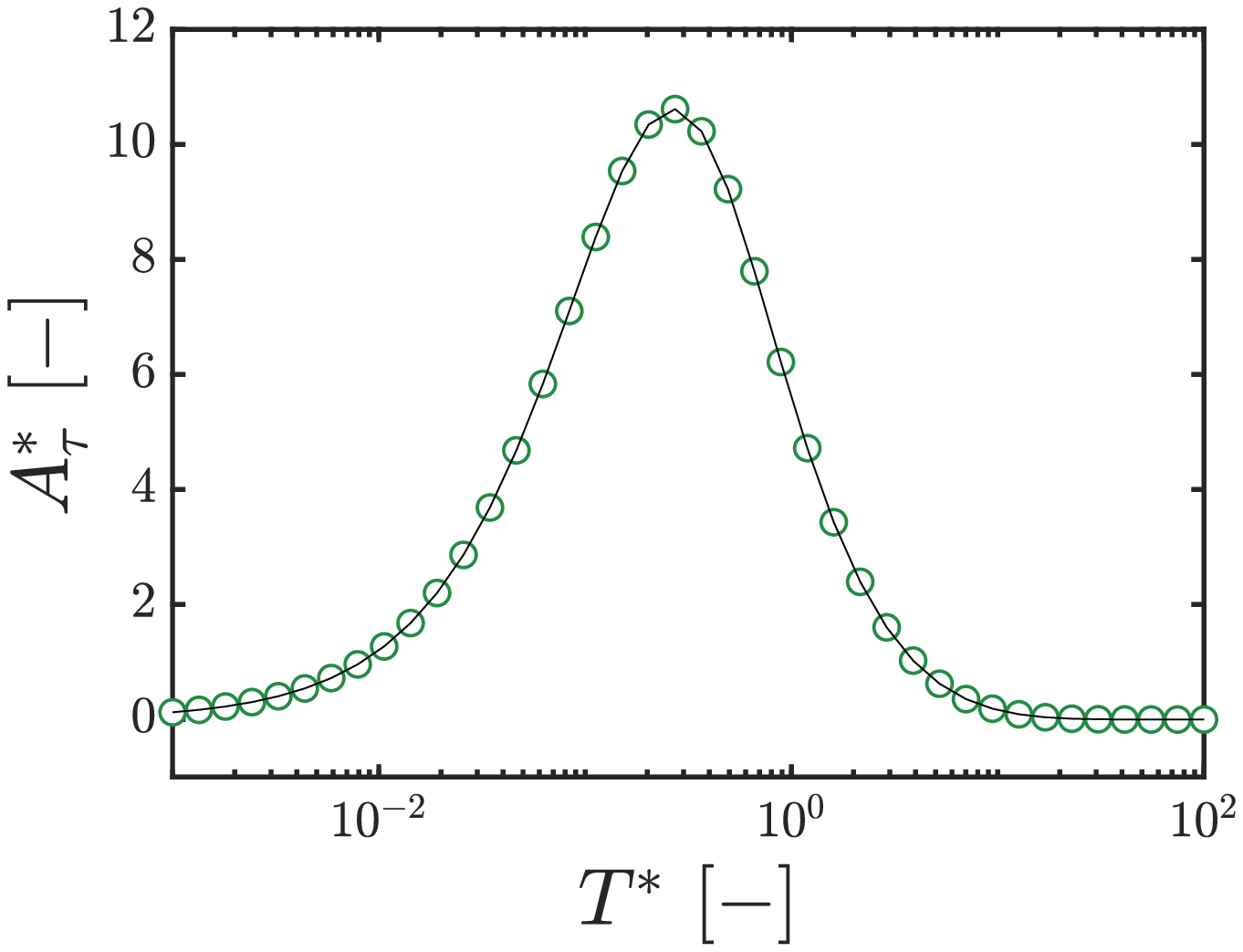}}
	\subfigure[]{
		\label{Fig.antithixotropy_hysteresis_area}
		\includegraphics[width=0.45\textwidth]{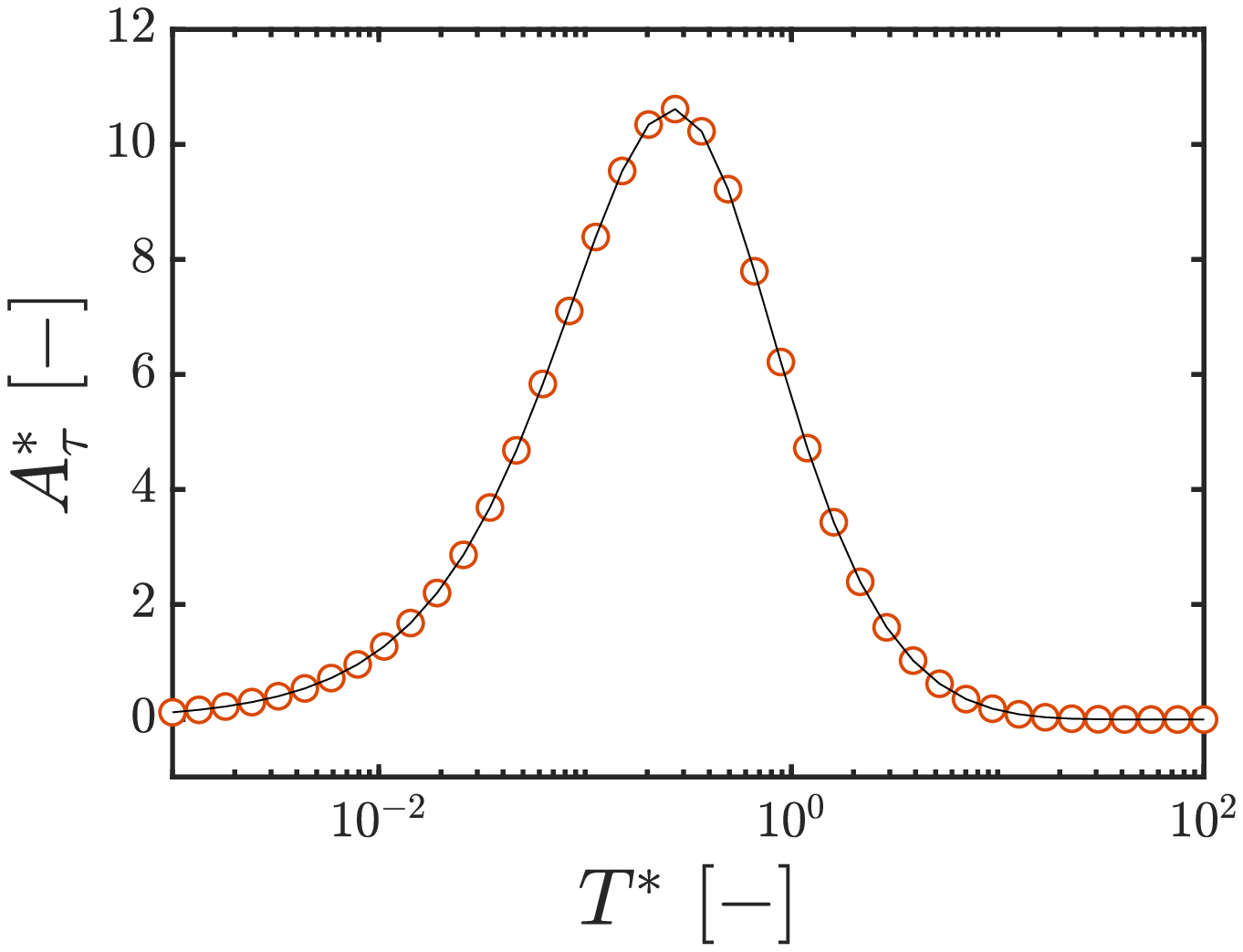}}
	\hspace{0.01\linewidth}
		\subfigure[]{
		\label{Fig.Giesekus_LVE_hysteresis_area}
		\includegraphics[width=0.45\textwidth]{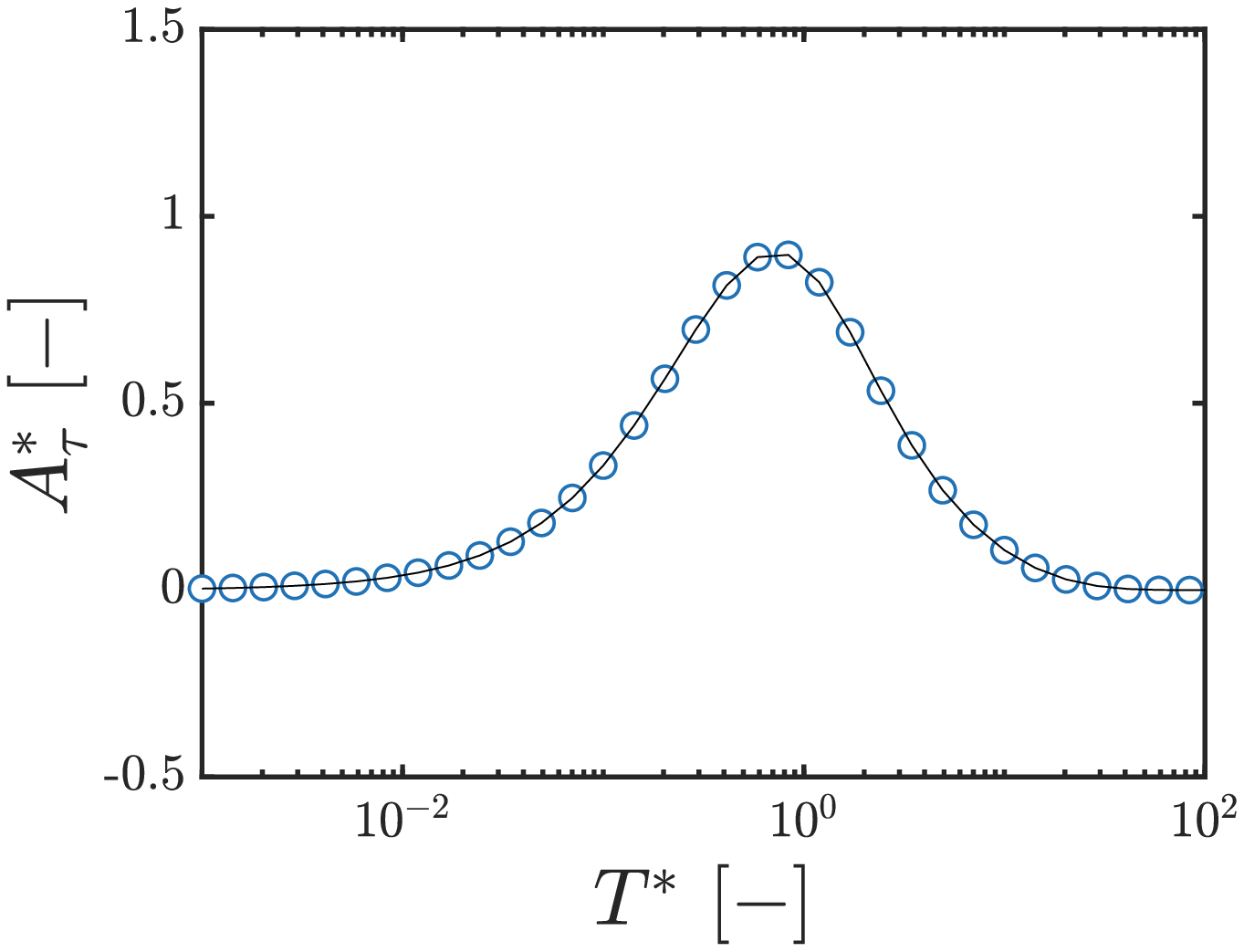}}
	\hspace{0.01\linewidth}
		\subfigure[]{
		\label{Fig.Giesekus_NLVE_hysteresis_area}
		\includegraphics[width=0.45\textwidth]{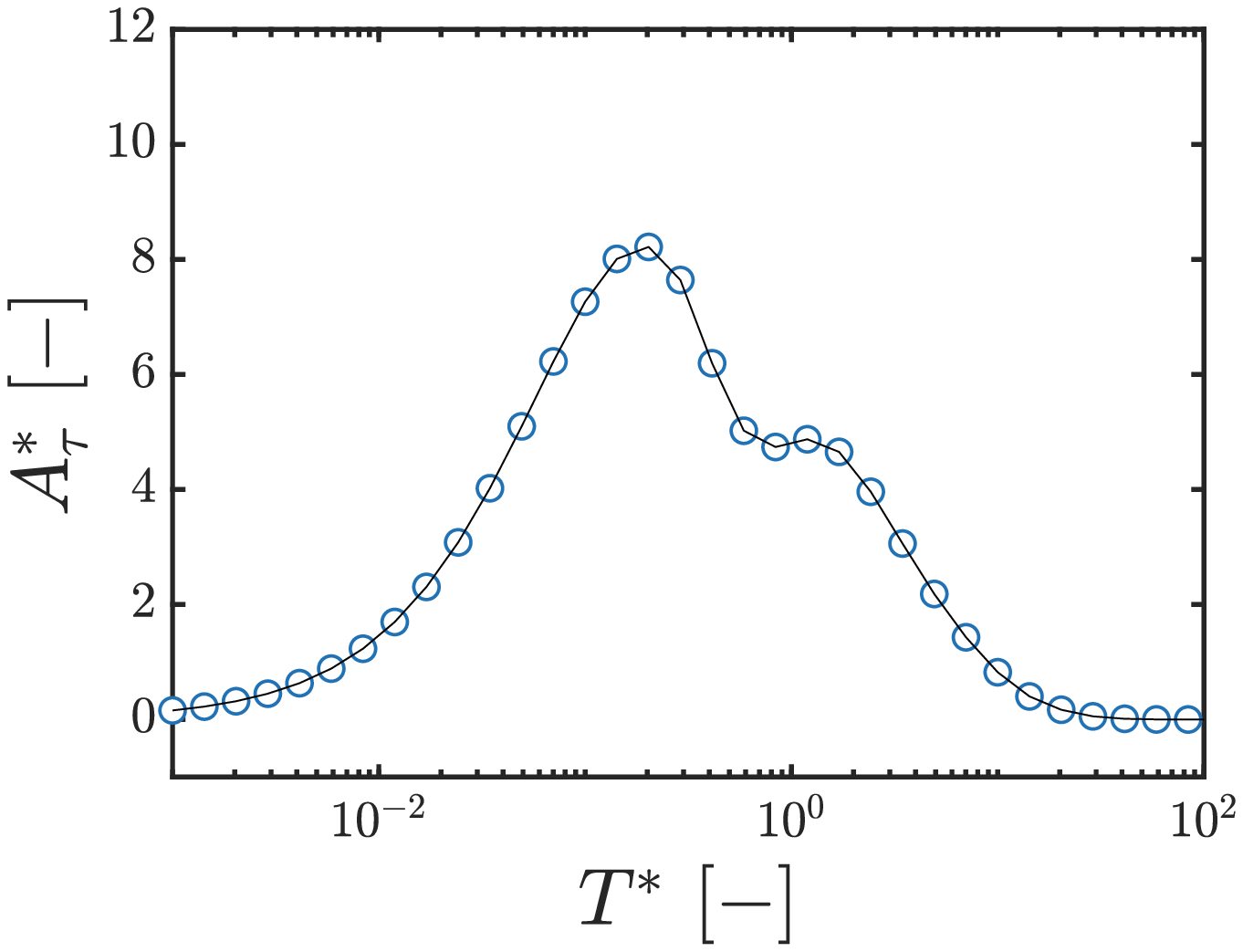}}
	\hspace{0.01\linewidth}
	\caption{The dimensionless area of the hysteresis loop, $A^*_\tau$, as a function of ramping time per decade, $T^*$, for (a) the thixotropic model, (b) the anti-thixotropic model, (c) the Giesekus model at linear ramping, and (d) Giesekus model at nonlinear ramping.}
	\label{Fig.area}
\end{figure}

Fig.~\ref{Fig.area} shows $A^*_\tau$ as a function of $T^*$ for hysteresis loops of the three models calculated in Fig.~\ref{Fig.comparison_model}. For the thixotropic model, a bell-shaped curve is observed, shown in Fig.~\ref{Fig.thixotropy_hysteresis_area}. This is consistent with results observed in a couple of thixotropic materials \cite{Divoux2013}, and the time $T^*$ at which the area is maximized suggests a characteristic timescale for the microstructure dynamics. In this study, the maximum is found at $T^*$ around 0.27. Remember that the $T^*$ defined in this study uses only $k_A$ and is therefore an underestimate of the true or average $ \mathcal{T}_+ $ (Eq.~\ref{Eq.thixo_define_timescale}). It should be noted that the hysteresis area and its maximization do not depend on the number of shear rates per decade, $n$, so long as $n$ is large enough to approximate the continuous ramping, which is the case of this study. But they may depend on the range of shear rates, which is kept as constant (10 to 0.1) here. Studying the effect of shear rates is outside the scope of this study. Fig.~\ref{Fig.antithixotropy_hysteresis_area} shows $A_\tau^*$ for the anti-thixotropic model, which, similar to that for thixotropy, displays a bell shape with maximum area found at $T^* \approx 0.27$.

Fig.~\ref{Fig.Giesekus_LVE_hysteresis_area} and \ref{Fig.Giesekus_NLVE_hysteresis_area} are $A_\tau^*$ at different $T^*$ for linear, small Wi ramping ($\dot \gamma^*$ decreases from 0.1 to 0.001 and then increases to 0.1 again) and nonlinear, large Wi ramping ($\dot \gamma^*$ changes between 10 and 0.1). For linear ramping, consistent with observation in Sec.~\ref{sec:results_model_hysteresis}, there is no self-intersection and the hysteresis area first increases then decreases with the ramping time, and $A_\tau^*$ shows a bell shape with a single maximum. The hysteresis area maximization is found at approximately $T^*=1$, corresponding to the viscoelastic Deborah number $\rm De \approx 1$. The shape of $A_\tau^*$ is more complicated for the nonlinear case,  which shows two peaks, with one peak found at dimensionless ramping time around 1, another lower than 1. The appearance of multiple maxima suggests a change in the direction of hysteresis.

The dependence of hysteresis area $A_\tau^*$ on the ramping time $T^*$ provides another distinction between the three dynamics, where thixotropy and anti-thixotropy have only one maxima, whereas viscoelasticity can have more than one maxima, depending on the range of shear rates. On the other hand, a characteristic timescale can be found at which the hysteresis area is maximized, which allows the comparison of characteristic timescales for different models and materials. However, the thixotropic timescale found in this way is protocol dependent: based on different ranges of shear rate applied, numbers of point tested, and so on, different values are expected. Moreover, only one timescale can be obtained in this way, whereas real systems usually demonstrate dynamics in a distribution of timescales, which can be obtained by spectrum analysis \cite{Sen2022}.

\section{\label{sec:results_exp}Experimental Results}

To experimentally demonstrate the distinguishing features of hysteresis loops identified in different models, in this section, we test three different materials, 3 wt$\%$ Laponite suspension, 8 wt$\%$ carbon black (CB) suspension, and 1 wt$\%$ 8M PEO solution, each representing one dominating dynamics, then we compare the measured hysteresis loops with the fingerprints shown in Sec.~\ref{sec:results_model}. 

\begin{figure}[h!]
	\centering
	\subfigure[]{
		\label{Fig.Laponite_stepshear}
		\includegraphics[width=0.3\textwidth]{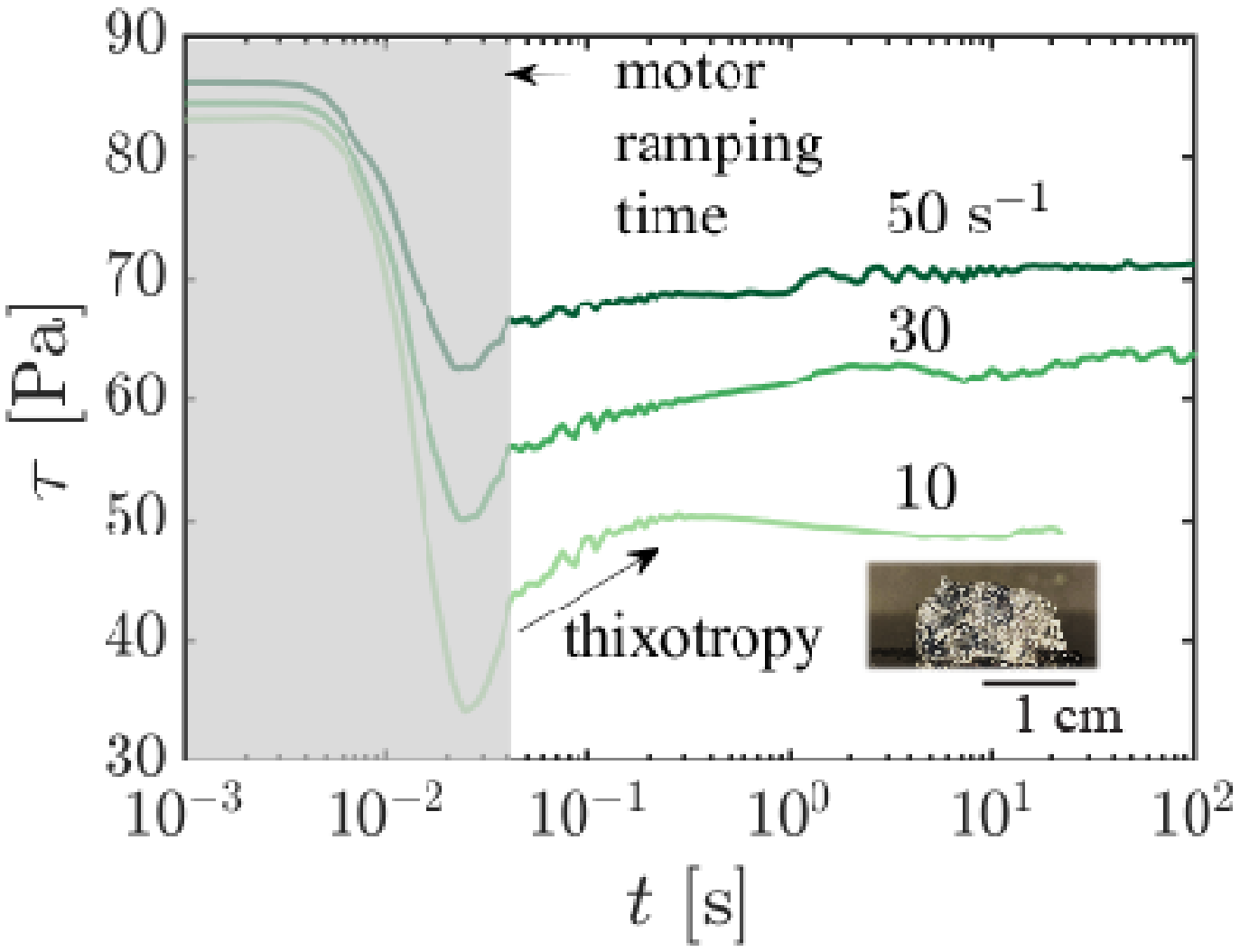}}
	\subfigure[]{
		\label{Fig.CB_stepshear}
		\includegraphics[width=0.3\textwidth]{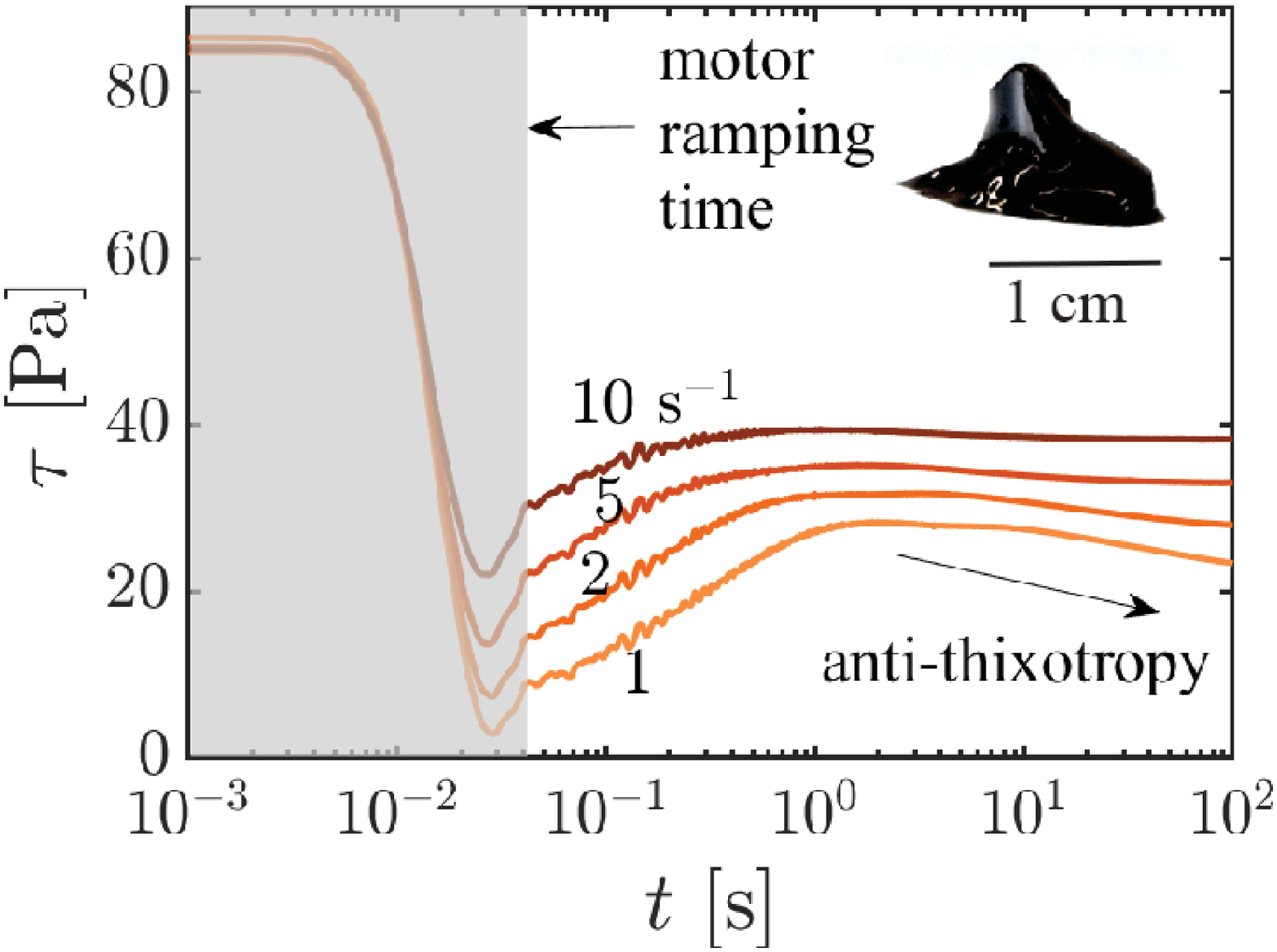}}
	\hspace{0.01\linewidth}
		\subfigure[]{
		\label{Fig.PEO_stepshear}
		\includegraphics[width=0.28\textwidth]{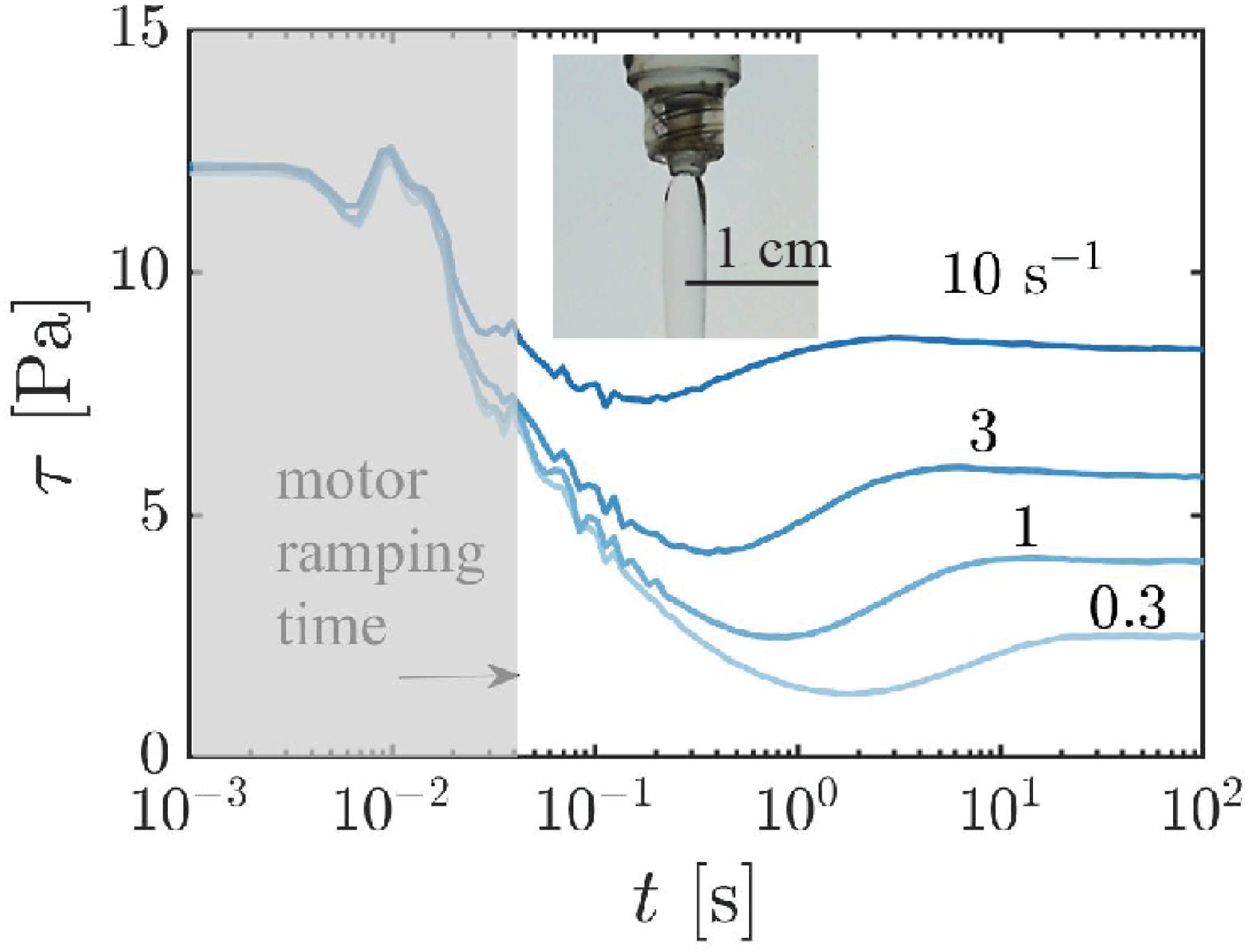}}
	\hspace{0.01\linewidth}
	\caption{Photos and transient stress responses in step-down in shear rate tests for (a) thixotropic 3 wt$\%$ Laponite suspension, (b) anti-thixotropic 8 wt$\%$ carbon black suspension, and (c) viscoelastic 1 wt$\%$ 8M PEO solution.}
	\label{Fig.material}
\end{figure}

Aqueous Laponite is a soft gel with thixotropy. Laponite is a synthetic disc-shaped crystalline colloid composed of hydrous sodium lithium magnesium silicate ($\rm Na^+_{0.7}[Si_8Mg_{5.5}Li_{0.3}O_{20}(OH)_4]^-_{0.7}$) \cite{Bonn1999}. The single disc-shaped Laponite crystals are typically 25 nm in diameter and 0.92 nm in thickness \cite{Cummins2007}. Because of the chemical structure, the faces of the disks are charged negatively and the sides of the disks can be charged positively, when the particles are suspended in an aqueous solution. This results in an electrostatic attraction between the faces and the sides of the disks. Therefore, they come together and form a "house of cards" structure. The structure of Laponite breaks down under high shear and builds up again when the applied shear rate is decreased/removed, resulting in thixotropy. The Laponite suspension used in this study is flowable but with finite yield stress, and therefore is dramatically shear-thinning, as shown in the inset photo in Fig.~\ref{Fig.material}\textcolor{blue}{(a).} One can estimate the yield stress based on the photograph \cite{Hossain2022}, which is around 43~Pa. The detailed calculation using this protorheology \cite{Hossain2022, Hossain2022a} test is shown in Supporting Information. As shown in Fig.~\ref{Fig.material}\textcolor{blue}{(a)}, when the applied shear rate is decreased from 100~$\rm s^{-1}$ to 50, 30, and 10~$\rm s^{-1}$ separately, shear stress increases with time.

Anti-thixotropy is less frequently observed but is present in some CB suspensions \cite{Wang2022, Hipp2019}. Fig.~\ref{Fig.material}\textcolor{blue}{(b)} shows a photo of 8~wt$\%$ CB suspension (inset), which demonstrates this is a yield stress fluid with yield stress larger than 68 Pa, estimated using protorheology \cite{Hossain2022}, see the Supporting Information for detailed calculation. The transient stress responses of CB in mineral oil suspension under a step down in shear rate test is shown in Fig.~\ref{Fig.material}\textcolor{blue}{(b)}. The CB suspension demonstrates a short time thixotropy and long time anti-thixotropy, as shown in our prior work \cite{Wang2022}, as the shear stress first recovers then decays. CB is virtually pure elemental carbon in the form of fractal particles \cite{Usersbook}. It exhibits a hierarchy of morphological features: the fundamental building blocks of CB, the primary particles, are strongly fused by covalent bonds into fractal aggregates, and the individual aggregates join together by van der Waals forces to form agglomerates. When the shear rate is decreased, the large but loosely-connected agglomerates are formed due to the attraction between them, which increases the viscosity of the suspension, and therefore, the shear stress recovers at short time, demonstrating thixotropic dynamics. While at long times, those fractal agglomerates self-organize, interpenetrate, and densify, resulting in a lower hydrodynamic volume and viscosity, therefore, the stress decreases, showing anti-thixotropic dynamics \cite{Wang2022}. The dynamics of CB suspension is dominated by anti-thixotropy at timescales longer than 1~s. 
 
Fig.~\ref{Fig.material}\textcolor{blue}{(c)} shows the shear stress decay function for aqueous PEO solution. PEO is a flexible, non-ionic water-soluble polymer \cite{Ebagninin2009}, which is known to be viscoelastic. The nonlinear viscoelasticity of PEO solution, specifically the elastic normal stress in shear, is demonstrated by the die swell shown in the protorheology photo in Fig.~\ref{Fig.material}\textcolor{blue}{(c)}. Under a step down in shear rate flow, when the applied shear rate is decreased from 30~s$^{-1}$ to 10, 3, 1, and 0.3~s$^{-1}$ separately, the stress of PEO solution shows a decay with time as a result of viscoelastic stress relaxation. Before reaching the steady state, a stress undershoot is present because of nonlinearity.

\begin{figure}[h!]
	\centering
	\includegraphics[width=0.8\textwidth]{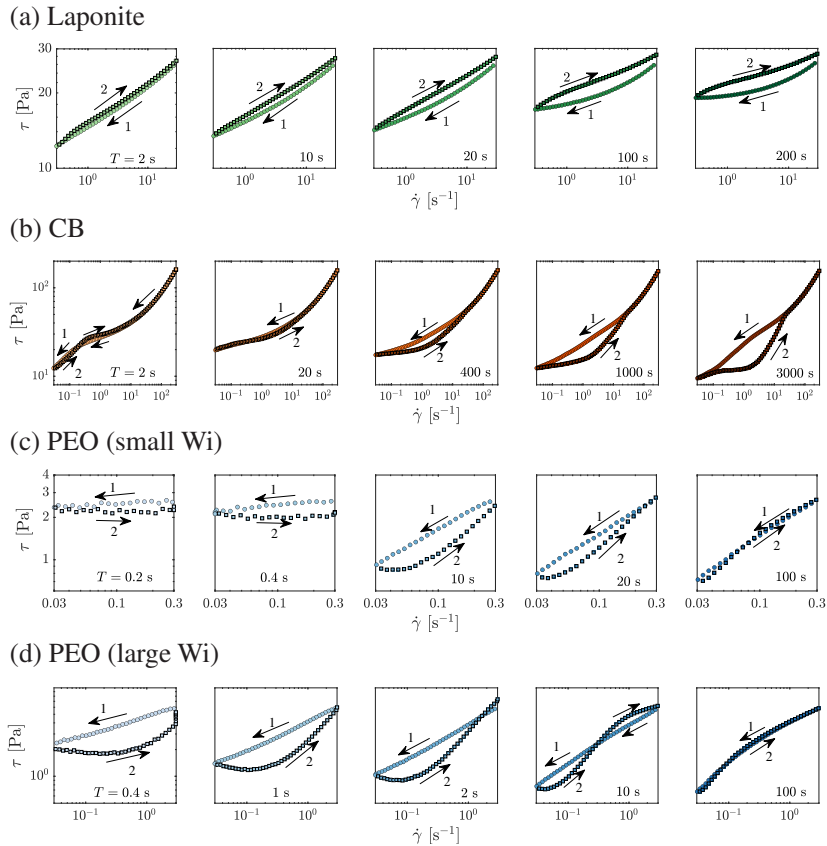}
	\hspace{0.01\linewidth}
	\caption{Measured hysteresis loops for (a) the thixotropic Laponite suspension, (b) the anti-thixotropic carbon black suspension, (c) the viscoelastic PEO solution at small Wi, and (d) the viscoelastic PEO solution at large Wi. From left to right, the ramping time increases, and the ramping rate decreases.}
	\label{Fig.experiment_hysteresis}
\end{figure}

The hysteresis loops for the thixotropic Laponite suspension, anti-thixotropic CB suspension, viscoelastic PEO solution in small and large Wi are measured at different ramping rates per decade, $T$, as shown in Fig.~\ref{Fig.experiment_hysteresis}. For thixotropic Laponite, $T$ ranges from 2 to 200 s, and the results are shown in Fig.~\ref{Fig.experiment_hysteresis}\textcolor{blue}{(a)}, where at very quick ramping ($T=2~\rm s$), the downward and upward ramping curves overlap and show a shear-thinning behavior. The hysteresis loops, when observable, are clockwise. As $T$ gets larger, the area starts to decrease, but never reaches zero. This is because Laponite is known to show long-time aging, where the properties keep changing without saturation.

Anti-thixotropic hysteresis loops for the CB suspension with $T$ ranging from 2 to 3,000~s, as shown in Fig.~\ref{Fig.experiment_hysteresis}\textcolor{blue}{(b)}, are counter-clockwise. We see from the step-down in shear rate tests that CB suspension shows short-time thixotropy, when the timescale is shorter than around 1~s. This is also reflected in hysteresis loops, when the ramping time per decade $T = $ 2~s, two self-intersections can be found, where the direction of the loop changes from counter-clockwise to clockwise, consistent with the thixotropic hysteresis loops.

The hysteresis loops of viscoelastic PEO solution are shown in Fig.~\ref{Fig.experiment_hysteresis}\textcolor{blue}{(c)} ($\dot \gamma$ ranges from 0.3 to 0.03~$\rm s^{-1}$, linear regime, with $T$ ranging from 0.2 to 100 s) and \textcolor{blue}{(d)} ($\dot \gamma$ ranges from 3 to 0.03~$\rm s^{-1}$, nonlinear regime, with $T$ ranging from 0.4 to 100 s). From the plots, we can see that in the linear regime, where small shear rates are used, there is no self-intersection and the direction of the hysteresis loops is counter-clockwise. While in the nonlinear regime, self-intersection can appear and change the direction of the loop. In both cases, at very quick ramping, the downward and upward ramping curves are approaching a constant stress. This is more obvious in the linear case, while for the nonlinear case, the ramping time is approaching the experimental limit: the duration per step is 0.02~s, which is the motor ramping response time. Therefore, a shorter ramping time is required for the nonlinear case to show a stress plateau, which is outside the instrumental limit in this study.

The distinguishing features of the hysteresis loops for each material in Fig.~\ref{Fig.experiment_hysteresis} are summarized in Table~\ref{tab:features}. For Laponite, the direction of the hysteresis loops is clockwise, at very quick ramping it shows minimal shear-thinning, the ramping down hysteresis curves are lower than the steady state flow curve, and there is no self-intersection. These fingerprints suggests that the dominating dynamics in Laponite is thixotropy, which is consistent with the step-down in shear rate test results (Fig.~\ref{Fig.Laponite_stepshear}) and the microstructure of Laponite suspensions. For the CB suspension, for most ramping times, $T$, the direction of the hysteresis loops is counter-clockwise, it shows shear-thinning at very quick ramping, the ramping down hysteresis curves are higher than the steady state flow curve, and there are self-intersections at short ramping times. These fingerprints suggests that the dominating dynamics of CB is anti-thixotropy, though thixotropy appears at short times. For PEO solution at low shear rates (linear regime), the direction of hysteresis loops is counter-clockwise, it shows a stress plateau in the low shear regime at quick ramping, the ramping down hysteresis curves are higher than the steady state flow curve, and there is no self-intersection in hysteresis loops at low rates. These features show that the dominating dynamics of PEO solution is viscoelasticity (elastic stress growth and relaxation), whereas the step shear rate tests of Fig.\ref{Fig.PEO_stepshear} may be ambiguous. Observing the features of hysteresis loops, we can fully differentiate the thixotropic, anti-thixotropic, and viscoelastic systems. 

\renewcommand{\arraystretch}{1.2}
\begin{table}[h!]
\caption{Features observed in experimentally measured hysteresis loops of suspensions of Laponite, CB, and PEO.}
\begin{tabular}{ p{4.5cm} p{3.8cm} p{3.8cm} p{3.8cm} } 
  \hline   \hline
Materials & Laponite & CB & PEO (small Wi) \\
  \hline
Direction of loop & Clockwise & Counter-clockwise & Counter-clockwise \\
Quick ramping & Minimal thinning & Minimal thinning & Stress plateau  \\
Ramping down stress  & Lower than ss$^*$  & Higher  &  Higher   \\
Self-intersection & No & Yes & No \\
\hline
Dominating dynamics & Thixotropy  & Anti-thixotropy & Viscoelasticity  \\
  \hline   \hline
\multicolumn{4}{l}{\small *ss represents steady state.} \\
\end{tabular}
\label{tab:features}
\end{table}

\section{\label{sec:conclusions}Conclusions}
We showed that hysteresis loops obtained by continuously ramping shear rate down-then-up is a promising way to distinguish thixotropy, anti-thixotropy, and viscoelasticity. We explored signatures of the most basic thixotropic, anti-thixotropic (which is introduced here by modifying the thixotropic kinetic model), and viscoelastic Giesekus models in hysteresis. For all three models, the area and shape of hysteresis loops depend on the ramping time per decade, $T^*$. At both short and long ramping times, $T^*$, the downward and upward ramping curves overlap and there is no hysteresis loop observed. The hysteresis can only be observed at intermediate $T^*$. Several fingerprints of hysteresis loops for different models are found, which can be used to differentiate the three dynamics. The first feature is the direction of hysteresis loops, clockwise for thixotropy, but counter-clockwise for anti-thixotropy and viscoelasticity. The second feature is found at quick ramping at small $T^*$, minimal shear-thinning is observed for thixotropy and anti-thixotropy, but a stress-plateau in the low shear rate regime appears for viscoelasticity. It should be noted that the hysteresis loops for viscoelastic models show a dependence on the minimum and maximum shear rates, and the distinguishing features become more complicated for nonlinear viscoelasticity. 

Additionally, we experimentally tested the hysteresis loops for there different materials at different $T$, the Laponite suspension, CB suspension, and PEO solution. From the characteristic features of the hysteresis loops, we confirm that the dominating dynamics at these test conditions of Laponite is thixotropy, of CB is anti-thixotropy, and of PEO solution is viscoelasticity. This conclusion is consistent with the step-down in shear rate test results, and the microstructure of the three materials, showing that fingerprints of hysteresis loops can be a credible criterion to distinguish whether time-dependent dynamics are dominated by thixotropy, anti-thixotropy, and viscoelasticity. Of course, all three dynamics can occur in the same system over different timescales, e.g., CB suspension has been proved to show short-time thixotropy and long-time anti-thixotropy \cite{Wang2022}, and most thixo-viscoelastic models are constructed by coupling a viscoelastic model with thixotropic kinetics, therefore showing both dynamics at different timescales \cite{Blackwell2014}. Our features observed here can help identify the dominating dynamics of systems at the timescale of interest.

The representative features of the thixotropy, anti-thixotropy, and viscoelasticity in hysteresis are observed and compared by considering the most fundamental thixotropic and anti-thixotropic models and the Giesekus model. However, there are numerous more complicated models that have not been considered in this paper, the hysteresis loops of which depending on the specific model forms and parameter values. Future studies should explore other models, such as more advanced thixo-elasto-visco-plastic models (TEVP) \cite{Ewoldt2017, Varchanis2019} in hysteresis. Besides, irreversible aging, which can also generate a hysteresis loop \cite{Mewis2009}, is not considered in this paper. We believe studying the fingerprints of irreversible aging can offer additional insight into the distinction between irreversible aging and thixotropy, but this is out of scope of this paper. Other non-ideal effects, such as viscous heating, shear banding, fluid inertia, and heterogeneous shear rate distributions in the sample are likely to happen during hysteresis \cite{Mewis2009, Divoux2013}, which are not discussed in this paper. Studying those non-ideal effects, and how they influence the shapes and fingerprints can benefit future research on using hysteresis to rule out the experimental artifacts during experiments \cite{Ewoldt2015}. 

The results presented here demonstrate that rheological hysteresis, which is frequently used to demonstrate thixotropy, can also appear in viscoelastic and anti-thixotropic materials. Great caution needs to be exercised to study thixotropic, anti-thixotropic, and viscoelastic dynamics in hysteresis. On the other hand, hysteresis loops for thixotropic, anti-thixotropic, and viscoelastic fluids have different fingerprints, which can be used as a credible protocol to help differentiate between the three dynamics. The results are important for thixotropic, anti-thixotropic, and viscoelastic systems, but also have potential to be used in studying aging, viscous heating, and shear banding and other rheological phenomena. 
\newpage

\bibliographystyle{unsrt}
\bibliography{hysteresis}

\end{document}